\documentclass[aps,superscriptaddress,preprint,showpacs,nofootinbib]{revtex4}
\usepackage{bm}
\usepackage{dcolumn}
\usepackage{epsfig}
\begin{document}

\title{Short-range three-nucleon forces and low-energy constants}

\author{J.~Adam, Jr.}
  \affiliation{ 
    Nuclear Physics Institute, \v{R}e\v{z} near Prague, 
    CZ-25068, Czech Republic}   
%   \email{adam@ujf.cas.cz}
\author{M.T.~Pe\~na}
  \affiliation{ 
    Instituto Superior T\'ecnico, Centro de F\'{\i}sica das Interac\c c\~oes Fundamentais,
    and Department of Physics, Av. Rovisco Pais, P-1049-001 Lisboa, Portugal }
\author{A.~Stadler}
  \affiliation{ 
    Centro de F\'{\i}sica Nuclear da Universidade de
    Lisboa, Av. Gama Pinto 2, P-1649-003 Lisboa, Portugal} 
  \affiliation{ 
    Departamento de F\'\i sica, Universidade de \'Evora,
    Col\'egio Lu\'\i s Verney, P-7000-671 \'Evora, Portugal}

\date{\today}

\begin{abstract}
The $3N$ forces due to $\pi-\rho, \pi-\sigma$  and $\pi-\omega$
exchanges following from the nucleon Born diagrams and diagrams with an
intermediate $N^{*}(1440)$ are re-analyzed. The cancellation between
$\pi-\sigma$  and $\pi-\omega$ forces is rather sensitive to the values
of the coupling constants and to the form of the $\pi NN$ vertex.
Experimental uncertainties in the parameters of the TM $\pi-\pi$ potential
are assessed. They lead to uncertainties in theoretical predictions of the
triton binding energy of about $\pm 0.4$ MeV. The low-energy limit of
 $\pi-\sigma$  and $\pi-\omega$ potentials is performed. It defines the coupling
constants of effective contact $\pi NNNN$ vertices, which are compared with
the corresponding contact vertices of Chiral Perturbation
Theory.
\end{abstract}
\pacs{21.30.-x, 21.30.Fe, 21.45.+v, 21.10.-k} \keywords{$3N$ forces, 
meson-exchange models, low-energy effective operators, low-energy 
constants, three-nucleon binding energy} 
%\preprint{nucl-th/03????}

\maketitle
%\setcounter{page} {0}  
%\vspace{1cm}
%==========================================================
\section{Introduction}

Nucleon-nucleon ($NN$) and three-nucleon ($3N$) potentials to be used 
in calculations of properties of few nucleon systems at low and 
intermediate energies are these days successfully modeled through meson 
exchanges. Modern $NN$ potentials based on this picture provide an 
impressive description (with $\chi^2$/datum $\sim$ 1) of the $NN$ 
scattering  data below the pion production threshold and of the 
deuteron properties \cite{CD-Bonn,Nijmegen}. These potentials often 
employ the underlying meson-nucleon dynamics only in one-boson-exchange 
approximation. One reason for this is clearly the desire for a 
relatively simple formalism, but it was also supported by extensive 
studies of the Bonn group \cite{Bonn} which showed numerous 
cancellations between classes of diagrams with two and more exchanged 
mesons. The meson-exchange approach with heavy  meson (and nucleon) 
resonances has been supplemented recently by a model-independent 
construction from a low-energy realization of QCD: chiral perturbation 
theory (ChPT). While some theoretical uncertainties in the formulation 
of this theory persist \cite{Beane}, the construction of the $NN$ 
potential in this framework is already approaching maturity 
\cite{Epelbaum,Epelbaumnew,Machleidt} with the  next-to-next-to leading 
order (NNLO) potential also describing the $NN$ data fairly well. 
Moreover, a recent paper \cite{saturation} by the J\"ulich-Bochum group 
provides a detailed comparison of ChPT with phenomenological potentials 
based on heavy meson exchange, and shows that the low-energy constants 
of ChPT appearing in the two-nucleon sector of the chiral Lagrangian 
agree reasonably well with those following from reducing the heavy 
meson operators to their contact form. Therefore a mutual positive 
influence between  the  two approaches should be expected for the 
future: the phenomenological potentials should provide a hint on the 
values  of the chiral low-energy constants (which are not always easily 
obtained from the experiment) and the chiral constraints should be 
imposed  on the phenomenological models at low energies. In this spirit 
some studies of ChPT $3N$  force were performed 
\cite{HuberFBS,Epelbaum3N}, although more extensive numerical 
calculations are still needed. To describe the data one might have to 
consider  the chiral $3N$ force from the next order of the chiral 
expansion \cite{Epelbaum3N}. 

\begin{figure}
\epsfig{file=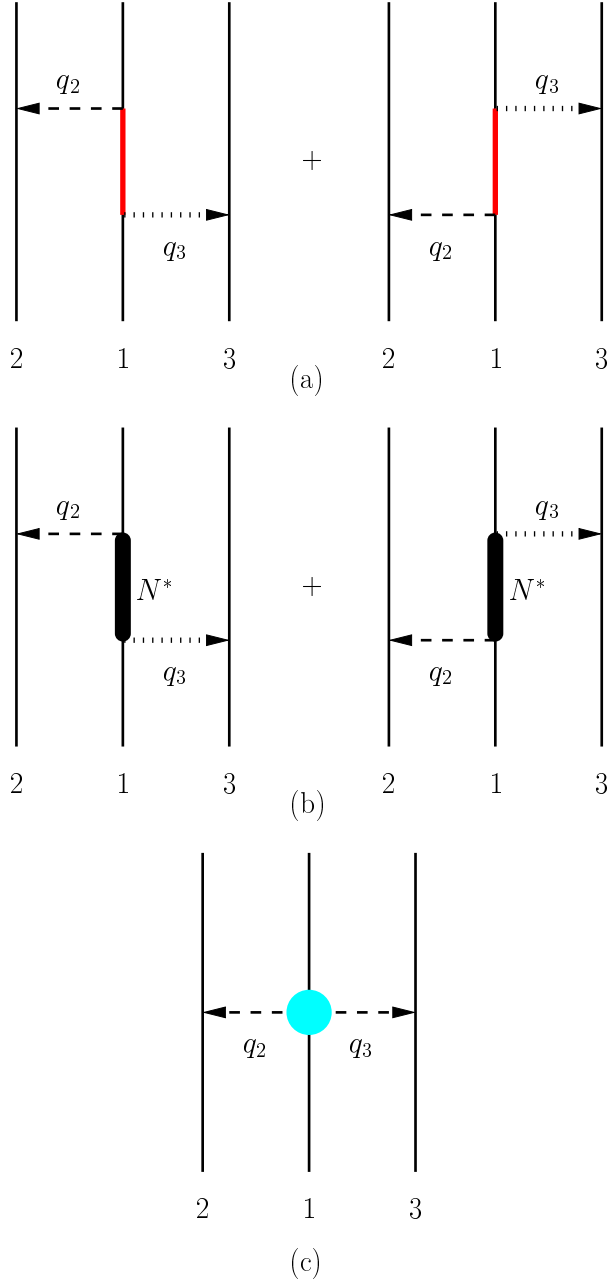,width=8cm} 
 \caption{Contributions to the
$3N$ force. Diagrams for (a) nucleon Born contributions, (b)
intermediate Roper resonance, and (c) $\pi$ re-scattering. Diagrams
(a) and (b) involve one pion and one heavy-meson exchange. They are
called ``long-short range $3N$ forces''. From diagrams (a), iterations of
the $NN$ potential are removed, as described in Appendix \ref{A:Born}.
Diagram (c) involves only pion exchanges. It is called a ``long-long range
$3N$ force''.  The meson momenta are defined as ${\bf q}_i= {\bf
p}_i^\prime- {\bf p}_i, \, i=1,2,3$.} 
\label{F:short-longandlong-long} 
\end{figure}
 
In the present paper we investigate the role of some short-range heavy 
meson exchanges in the $3N$ potential with respect to the binding 
energy of the $3N$ system. In addition to the dominant $\pi-\pi$ 
component of the $3N$ potential \cite{TM,TM1}, we include the force 
following from $\pi-\rho$ exchange \cite{Ellis} and the terms involving 
$\pi-\omega$ or  $\pi-\sigma$   exchanges from Ref.\ \cite{CPR}. The 
latter potentials were obtained both from diagrams with the Roper 
resonance $N^{*}(1440)$ (Fig.\ \ref{F:short-longandlong-long}b) and 
from the nucleon Born terms (often imprecisely called pair- or 
Z-diagrams) where an off-mass-shell nucleon propagates in the 
intermediate state (Fig.\ \ref{F:short-longandlong-long}a). All 
contributions of the Born diagrams are of relativistic order. The parts 
with negative energy propagation, being genuine ``Z-diagram''  
contributions, are included in the irreducible $3N$ potential. Also the 
part with  off-shell nucleons propagating with positive energy cannot 
be omitted. These terms differ in general from the iteration of the 
$NN$ potential (which is not to be included in the $3N$ force and has 
to be subtracted from the total amplitude explicitly) and thus also 
contribute to the $3N$ potential.  

The purpose of this paper is to study the effects of the $3N$ forces 
based on pion-pion and pion-heavy-meson exchanges. It is discussed how 
strongly they depend on experimental uncertainties in the determination 
of the phenomenological constants in terms of which these potentials 
are parameterized, and on the model dependence of the nucleon Born 
contributions. These inherent uncertainties should provide -- for such 
semi-phenomenological meson exchange models -- theoretical error bars 
of the calculated bound state energy of the trinucleon system. 

The latest version of the phenomenological Tucson-Melbourne (TM) 
potential \cite{TM,TM1} $\pi-\pi$ potential is parameterized in terms 
of three constants $a,b,d$. These constants are determined from $\pi N$ 
scattering data and we look at the variations of the triton binding 
energy within the range of their experimental uncertainty, 
complementing thereby the usually studied sensitivity to the variation 
of the form factor cut-off parameter.  

The pion-heavy-meson exchanges contributions contain also some not very
well determined constants, in particular the coupling constants in 
vertices involving the Roper resonance $N^*$. But besides that the Born 
terms also depend on the employed form of the $\pi NN$ couplings. It is 
often assumed that the pseudovector (PV) coupling is preferable: this 
is based on the experience with modeling relativistic OPE $NN$ 
potentials where the pseudoscalar (PS) $\pi NN$ coupling implies 
unrealistic enhancement of the intermediate negative energy states. It 
is also claimed that a PV coupling is preferred because it does not 
require the non-minimal contact terms to conform with chiral symmetry 
in processes like pion electroproduction on the nucleon. 

However, in constructing the $\pi N$  scattering amplitude, or 
amplitudes of heavy meson production in pion  absorption on the 
nucleon, one has to be more careful. In these cases  one cannot state 
that the use of PV coupling minimizes the contact  Lagrangians: while 
it is true for the isospin-even $\bar{\psi} \psi \Phi^2$  vertex, the 
isospin-odd contact Weinberg-Tomozawa interaction is  actually close to 
zero in the representation with PS coupling. If one tries  to include 
the heavy mesons the situation becomes even less clear: to  include 
them one has to extend the symmetry in a model dependent way, e.g. to 
require the {\em local} SU(2) $\times$ SU(2) symmetry  dynamically 
realized (and broken) either in the Young-Mills fashion or  using the 
concept of hidden symmetry realization. The construction of  these 
(approximately) chiral symmetric Lagrangians including heavy  mesons is 
thus model dependent. It does not allow to formulate the  consistent 
chiral counting scheme and even at the tree level the  dynamical 
content has not been sufficiently constrained  by detailed analysis of 
wide range of hadronic processes, which they in  principle should 
describe. 

In this paper we therefore consider only  the Born amplitudes and the 
amplitudes with intermediate excitation of the Roper $N^*$(1440) 
resonance,  which have been proposed in Ref.\ \cite{CPR}, and have 
never been  included in realistic calculations of trinucleon bound 
states. Since we conclude from the reasoning above that there is no 
strong reason to  prefer the Born contributions obtained with PV $\pi 
NN$ coupling, we calculate also their PS version. In contrast with Ref. 
\cite{CPR} we include all these potentials in  $3N$ Faddeev 
calculations exactly, i.e., without using perturbation  theory. Their 
contributions to the binding energy are given individually. 

Finally, we relate some of these short-range $3N$ potentials to the 
corresponding  counterterms from the chiral Lagrangian. We have deduced 
from the potentials with heavy meson exchange the effective contact 
low-energy four-nucleon--pion  coupling constants and attempted to 
relate them with the constants of NLO interactions of  ChPT 
\cite{Epelbaum3N,FriarKolck}. When the heavy meson propagator is 
reduced  to a point, it appears that the pion--heavy-meson $3N$ 
potentials with PS $\pi NN$ coupling are closer to those obtained from 
the contact low-energy four-nucleon--pion NLO interactions of  ChPT 
\cite{Epelbaum3N,FriarKolck}  than their PV $\pi NN$ versions. 
Nevertheless, the comparison with  chiral low energy constants is not 
straightforward, since some of the effective interactions we obtained 
by taking the point limit are not included in the NLO interactions of 
ChPT, and can be transformed to that form only after certain 
approximation. This approximation however does not seem to be 
numerically supported by results of our model calculations. 

This paper is organized into four sections. Section II contains an 
overview on $3N$ forces. Section III presents the long-long and the 
long-short $3N$ forces and the numerical results. Section IV determines 
the low-energy constants and section V gives a summary and conclusions. 

\section{Brief overview of $3N$ forces}

Chiral symmetry has been recognized as an important guideline for the
construction of nuclear forces, into which the $\pi N$ amplitude enters 
as one of its building blocks, long before the advent of ChPT. The 
process of incorporating constraints from chiral symmetry (breaking) 
into theoretical studies of the two-nucleon interaction was pioneered  
by G.~Brown \cite{GBrown} in the early 1970's. Later, the same ideas 
were applied to the $3N$ force, for which two different approaches were 
developed in parallel: 

(i) One is based on the concept of partially conserved axial-vector 
current (PCAC) and the current algebra (CA) formalism, built from 
equal-time commutation relations for vector and axial-vector currents. 
It underlies the well-known Tucson-Melbourne (TM) two-pion-exchange 
$3N$ force \cite{TM,TM1} represented in Fig.\ 
\ref{F:short-longandlong-long}c. It goes beyond the static P-wave 
$\Delta$ contribution to the $\pi N$ amplitude, the only building block 
considered in its predecessor, the Fujita-Myazawa $3N$ force. Current 
algebra  and PCAC provide an elastic pion-nucleon scattering amplitude 
which includes the pion-nucleon $\sigma$-term, a direct measure of 
chiral symmetry breaking, which can be extracted from experiment. 

(ii) The other approach stems from the so-called ``effective chiral
Lagrangians'' for the $\pi N$ system. The first example of this 
approach is the Gell-Mann and Levy linear sigma model. Another example 
is the chiral Lagrangian with pseudo-vector $\pi N$ coupling, which 
underlies the construction of the so-called Brazilian $3N$ force in 
Ref.\ \cite{Robilotta}. More recently, ChPT, based on the Weinberg 
Lagrangian \cite{Weinberg} supplemented by multinucleon contact terms, 
became a systematic way to approach two and $3N$ forces \cite{Kolck}, 
and was applied to the description of low-energy hadronic physics. 

In particular, it was realized in Ref.\ \cite{FriarKolck} that although the
construction of the TM two-pion exchange $3N$ force employs  chiral
constraints for $\pi N$ scattering through CA, it missed further
constraints,  which arise from embedding that amplitude in the $3N$ system.
The conclusion was that the TM form of the two-pion-exchange force contained
a spurious term corresponding to contact terms  between two nucleons and
pions ($NN\pi \pi$). This spurious term can be generated also in ChPT with the
help of a pion field redefinition. But at the same time an additional
two-nucleon contact term arises ($NNNN\pi$), which has not been taken into
account by the TM group and which exactly cancels the first spurious contact
($NN\pi \pi$) contribution. As a net result, the so-called $c$ term of the TM
force should be dropped and the so-called $a$ term is modified. We present
here the effects  of the chirally imposed changes to the TM force on the
triton binding energy. A previous calculation of these effects \cite{CoonHan}
used a less realistic $NN$ interaction and a variational numerical method.

The CA program was extended to describe also $\pi N$-$\rho N$ transitions,
from which a $\pi-\rho$ exchange $3N$ force can be constructed. It seems
natural to include such a mechanism, given the important interplay between
$\pi$ and $\rho$ exchange in two-nucleon potentials. To model the $\pi-\rho$
$3N$ force, one can use  vector-meson dominance to access the $\rho$-analogue
of the off-mass-shell pion electroproduction. Chiral symmetry is in this case
supplemented by gauge invariance to constrain the Ward identity amplitude.  

The interest in the short-range $3N$ forces increased in mid 1990's,  
when it was found that the lack of binding energy of the triton is not 
the only experimental signature of a $3N$ force. According to 
\cite{FriarHuber} a spin-orbit structure of the $3N$ force, not present 
in the standard two-pion exchange based forces, may solve systematic 
deviations between the theoretical and the experimental maximum of the 
nucleon vector analyzing power $A_y(\theta)$ in elastic $nd$ scattering 
below 30 MeV. The TM or the Brazilian $3N$ force either produces no 
sizable effect or even worsens the disagreement with the data. But the 
study of H\"uber \textit{et al.} \cite{HuberFBS} indicated that the 
short-long range forces with a structure similar to that given by 
$\pi-\sigma$ and $\pi-\omega$ exchanges with an intermediate Roper 
resonance $N^*$ \cite{CPR} and suggested by the NLO ChPT 
\cite{FriarKolck} could potentially improve the description of $A_y$. 
Similarly, a purely phenomenological spin-orbit $3N$ force yields a 
noticeable improvement of the description of this observable 
\cite{Kievsky}. 

More recently, Meissner \textit{et al.}\ \cite{Meissner} actually 
claimed that the $A_y$ puzzle can be resolved by a chiral $NN$ force 
alone. However, the $NN$ interaction employed does not match the high 
quality of contemporary descriptions of $NN$ scattering data, in 
particular, for the crucial triplet P-waves 
\cite{FriarHuber,Machleidt}. The NLO chiral potential gives a 
\textit{qualitative} description of the phase shifts only for very 
small energies ($\leq 10$MeV), but for a truly quantitative fit one has 
to include NNLO corrections \cite{Machleidt,Epelbaumnew}. Entem and 
Machleidt showed that no low-$\chi^2$ $NN$ potential, neither one based 
on heavy-meson exchanges nor one constructed from ChPT, can solve the 
$A_y$ puzzle, since a good fit to the $NN$ phase shifts (in particular 
in the $^3P_J$ channel) does not allow the variation needed to fix this 
spin observable. In a recent paper of Epelbaum \textit{et al.}\ 
\cite{Epelbaum3N} it was found that NNLO $NN$ and leading order $3N$ 
chiral potentials still do not solve the $A_y$ problem completely. Also 
the Urbana $3N$ force (a Fujita-Myazawa force with a short-range 
regularization determined phenomenologically through a fit to the 
triton binding energy) does not provide a solution.  An interesting 
alternative was suggested by Canton \textit{et al.} \cite{Canton}, who 
constructed an additional $3N$ force by reducing $\pi-3N$ equations 
below the pion threshold. Their force is very similar in spin-isopsin 
structure to the $\pi-\sigma$ and $\pi-\omega$ force with intermediate 
Roper resonance discussed above, but it is multiplied by a part of an 
energy-dependent $NN$ T-matrix. 

In addition to the mentioned $nd$ scattering problem, a good description of
neutron-rich light nuclei is only possible \cite{Pieper} when a
phenomenological $3N$ interaction with three pion exchanges is added to the
original Urbana force \cite{Urbana}. Finally, let us mention that while
realistic $NN$ potentials alone predict  an equilibrium density for symmetric
nuclear matter that is too high, phenomenological $3N$ forces can change it
in the right direction \cite{matter}. 

This overall picture indicates that the long-range $\pi$-$\pi$
$3N$ force needs to be complemented by other physical mechanisms. 
 
%==============================================================================
\section{Long-Long and Long-short range $3N$ forces}

The phenomenology of the $NN$ interaction shows that the most important meson
exchanges  in every realistic OBEP are exchanges of the pion, of
the vector $\rho$ and $\omega$ mesons, and  of the (fictitious)
scalar $\sigma$ meson. Therefore, one should also investigate
the role of such exchanges in the modeling of a $3N$ force. 

\subsection{$\pi-\pi$ exchange force}
\label{S:pipi}

Given the importance of the one-pion-exchange potential (OPEP) in determining
the long-range part of the $NN$ interaction, it was natural that the first
$3N$ force considered was the two-pion exchange. The corresponding diagram is
shown in Fig.\ \ref{F:short-longandlong-long}c,
where the blob represents all possible pion re-scattering
processes. The iteration of the OPEP has to be subtracted from this Feynman 
diagram, since it is generated automatically by iterating the $NN$ potential
in the Faddeev equation. The $\pi N$ amplitude $t_{\pi N}$  entering the $3N$
force involves virtual pions. This off-mass-shell continuation is constrained
by the soft pion low-energy theorems. In this paper, we adopt the (modified)
TM version of $\pi-\pi$ force, which has been constructed to explicitly
conform with these requirements.  

The TM $\pi-\pi$ $3N$ force \cite{TM,TM1} was generated by applying the 
Ward identities of CA to the amplitude of axial-vector current 
scattering on the nucleon. Using PCAC, the resulting $\pi N$ amplitude, 
in the even and odd isospin channels, was written as an expansion in 
powers of the momentum transfer $t$ and the crossing-variable $\nu = 
(s-u)/4M$. Then the ``reality test'' was applied with the successful 
result that CA predicts the first four coefficients of the expansion in 
good agreement with the empirical subthreshold expansion coefficients 
obtained from dispersion relations. Therefore, it was possible to use 
accurate empirical $\pi N$ data to construct the nearly 
model-independent TM force. The dispersion analysis has been updated 
over time with the inclusion of new data, therefore  the coefficients 
of the TM force have also evolved.  

In terms of physical mechanisms, the non-Born part of the TM $\pi N$ 
amplitude contains terms that can be interpreted as $t$-channel 
$\sigma$ and $\rho$ exchanges, as well as contributions from the 
$\Delta$ resonance in the intermediate state. The contributions of the 
subtracted nucleon Born diagrams were analyzed in detail in 
\cite{piBorn,PenaCoon} and it was later shown numerically that they are 
relatively small \cite{Stadler}, at least as far as their contribution 
to the triton binding is concerned. 

As argued in \cite{FriarKolck}, the TM $\pi N$  amplitude, although
constructed from CA and obeying the chiral constraints in the $\pi N$
sector, should not be simply attached to two additional nucleon lines, 
since that would be inconsistent with the chiral counting for the $3N$ 
potential. More precisely, the TM $\pi N$  amplitude can be derived 
from the usual chiral Lagrangian after the pion  field re-definition 
\begin{equation}
  \bm{\pi^{\prime}}= \bm{\pi}\, (1- c)\, N^\dagger N \, ,
\label{redefin}
\end{equation}
where $c\approx \sigma/(m_\pi^2 f_\pi^2)$ is the constant of
the so-called TM $c$-term. The original chiral Lagrangian
in the ``natural'' representation acquires after such redefinition 
two additional terms (relevant up to the order considered)
\begin{equation}
  \Delta \mathcal{L}^{(1)}= -c\,  N^\dagger N\, \Bigrb\{
  ( \bm{\pi^\prime} \partial_\mu \partial^\mu \bm{\pi^\prime}
  + m_\pi^2 \bm{\pi^{\prime\, 2} )}+
  \frac{g_A}{2f_\pi} [ (\bm{\nabla}N^\dagger)\cdot \bm{\sigma}
  \bm{\tau\cdot\pi^\prime} N + 
  N^\dagger  \bm{\tau\cdot\pi^\prime} \bm{\sigma}\cdot (\bm{\nabla}N) ]
  \Biglb\} \, .
\label{Lpicont}  
\end{equation}

The TM amplitude includes the term corresponding to the first part of 
$\Delta \mathcal{L}^{(1)}$ (the $c$-term proportional to ${\bf q}_2^2+ 
{\bf q}^2_3$), but in the construction of the TM $3N$ potential the 
second term (which involves two nucleons) was not considered. 
Therefore, according to Ref.\ \cite{FriarKolck}, the TM $c$-term should 
be dropped and its $a$-term should be replaced by $a'= a-2 m_\pi^2\, 
c$. The TM $\pi-\pi$ force modified in this way was recently adopted in 
\cite{CoonHan} and labeled TM'(99). However, we would like to point out 
that it is not correct to transform away the whole $c$ term, because it 
includes  the contribution from the Born diagrams-- that contribution 
arises naturally also in  ChPT in its usual representation. 
Nevertheless, since this residual $c$-term is rather small and  the 
$\pi-\pi$ force is not the main subject of interest of this paper, we 
avoid introducing further redefinitions of the TM force and we adopt 
TM'(99) for our numerical calculations. 

Thus, the $\pi-\pi$ potential used in this paper reads 
% pi-pi TNI  
\begin{eqnarray}
  W_{\pi\pi}(1) &=&
 - (2\pi)^3  \frac{g^2}{4m^3}\,
 (\bm{\sigma}_2 \cdot {\bf q}_2)\, (\bm{\sigma}_3 \cdot {\bf q}_3)\,
  \tilde{\Delta}_\pi({\bf q}_2)  \tilde{\Delta}_\pi({\bf q}_3)
 \nonumber \\
& & \times
\left\{ ( \bm{\tau}_2 \cdot  \bm{\tau}_3 )
     \left[ a' - b\, {\bf q}_2 \cdot {\bf q}_3
            \right]
- (i \bm{\tau}_1 \cdot \bm{\tau}_2 \times \bm{\tau}_3)
d\, (i \bm{\sigma}_1 \cdot {\bf q}_2 \times {\bf q}_3) \right\}
   + 2 \leftrightarrow 3       \, , 
\label{Epipi}
\end{eqnarray}
where we use for the meson momenta ${\bf q}_i= {\bf p}_i^\prime- {\bf 
p}_i$, so that they are always pointing away from the ``active'' 
nucleon as shown in Fig.\ \ref{F:short-longandlong-long}c (to avoid 
confusion, we note that the different  convention with the first pion 
incoming and the second outgoing from the active line rescattering 
``blob'' is also often used in discussion of this potential). The 
overall momentum conservation is then given by ${\bf q}_1+ {\bf q}_2+ 
{\bf q}_3 =0$. The functions $\tilde{\Delta}_B({\bf q})$ are defined to 
contain the propagator function of the meson $B$ and the square of the 
strong $BNN$ form factor  $\tilde{\Delta}_B({\bf q})= F^2_{BNN}({\bf 
q}^2)/(m_B^2+ {\bf q}^2)$. The full potential contains additional four 
terms following from (\ref{Epipi}) by cyclic permutations. 

The parameters of the TM'(99) force are given by
\begin{eqnarray}
a' & = & \frac{\sigma}{f_\pi^2} -2m_{\pi}^2 
  \left( \frac{\sigma}{m_{\pi^+}^2 f_\pi^2} - \frac{g^2}{4m^3}+F'_{\pi NN}(0) 
  \frac{\sigma}{f_\pi^2} \right) \label{E:TMa}\\
b & = & - \frac{2}{m_{\pi^+}^2} 
   \left[ \bar{F}^+(0,m_{\pi^+}^2) - \frac{\sigma}{f_\pi^2} 
   \right] \label{E:TMb}\\
d & = & -\left[ \frac{\bar{B}^-(0,0)}{2m}+\frac{g^2}{4m^3} \right] \,.
\label{E:TMd}
\end{eqnarray}

Here, $\sigma$ is the pion-nucleon sigma term, $g$ is the $\pi NN$ 
coupling constant, $f_\pi$ the pion decay constant, $\bar{F}^+(\nu,t)$ 
and $\bar{B}^-(\nu,t)$ are isospin-even non-spin flip and isospin-odd 
spin-flip $t$-channel $\pi N$ amplitudes, respectively, with the 
nucleon pole term subtracted \cite{TM}. Note that we distinguish 
between the mass of the charged pions, $m_{\pi^+}=139.6$ MeV, and the 
isospin averaged pion mass $m_\pi= 138.0$ MeV. Although their 
difference is small, it affects the 3N force parameters in a noticeable 
way. The vertex form factor 
\begin{equation} 
F_{\pi NN}({\bf q}^2)=
 \frac{\Lambda_{\pi NN}^2-m_\pi^2}{\Lambda_{\pi NN}^2+{\bf q}^2} 
\end{equation} 
depends on the cut-off parameter $\Lambda_{\pi NN}$, which in the 
original TM force was taken to be $5.8$ $m_\pi$ in order to be consistent 
with a Goldberger-Treiman discrepancy of 3\% determined at the time. 
Based on more recent data, the Goldberger-Treiman discrepancy shrank to 
about 2\%, which corresponds to $\Lambda_{\pi NN}$ close to 7.1 
$m_\pi$. On the other hand, $\Lambda_{\pi NN}$ is frequently tuned to 
reproduce the triton binding energy in calculations that include only 
the $\pi-\pi$ exchange part of the $3N$ force. Such calculations 
require much lower values, close to 4 $m_\pi$. Clearly, the functional 
form of the form factor is not much constrained by the knowledge of the  
coupling constant  $g\, F_{\pi NN}(q^2)$ at the two points $q^2=0$ and 
$q^2=m_\pi^2$ and thus does not allow a strict determination of a 
cut-off mass. We keep therefore the by now traditional reference value 
$\Lambda_{\pi NN}=$ 5.8 $m_\pi$ as our standard one, but also vary it 
then between the limits indicated above to study the sensitivity of the 
results. In particular, we investigate the question whether the 
short-range forces are able to reduce the strong cut-off dependence, as 
one may  expect from arguments of ChPT. 

Adopting the values \cite{CoonHan} $g^2=172.1$, 
$\bar{F}^+(0,m_{\pi^+}^2)=-0.05$ $m_{\pi^+}^{-1}$, $\bar{B}^-(0,0)=8.6$  
$m_{\pi^+}^{-1}$, $\frac{\sigma}{f_\pi^2}= 1.40$ $m_{\pi^+}^{-1}$, and 
$\Lambda_{\pi NN}=5.8$ $m_\pi$, we obtain the coefficients given in Table 
\ref{T:TMconst}. Note that $a'$ depends weakly on $\Lambda_{\pi NN}$, a 
dependence often ignored in practical calculations that vary the 
cut-off mass. 
\begin{center}
\begin{table}[bt]
\begin{ruledtabular}
\begin{tabular}{cc}
$\Lambda_{\pi NN}/m_\pi$ \hspace{5mm} & $m_\pi a'$  \\
\hline
 4.1  &  -1.203 \\
 5.0  &  -1.154 \\
 5.8  &  -1.127 \\
 6.5  &  -1.112 \\
 7.1  &  -1.101 \\
\end{tabular}
\end{ruledtabular}
\caption{Expansion coefficient $a'$ of the $\pi$N amplitude used in the
TM $\pi-\pi$ force as a function of the $\pi NN$ vertex cut-off 
parameter $\Lambda_{\pi NN}$, in units of the isospin averaged pion 
mass. The other coefficients $b$ and $d$ do not depend on $\Lambda_{\pi 
NN}$ and have the values $m_\pi^{3} b=-2.801$ and  $m_\pi^{3} 
d=-0.754 $.} \label{T:TMconst} 
\end{table}
\end{center}
While the rather strong dependence of the $3N$ binding energy on 
$\Lambda_{\pi NN}$ has been recognized as a source of significant 
uncertainty, little attention has been paid to uncertainties 
originating from the experimental errors in the other parameters of the 
TM $\pi-\pi$ exchange force. Since the first publication of the TM 
force, the experimentally determined values of $g^2$, 
$\bar{F}^+(0,m_{\pi^+}^2)$, $\bar{B}^-(0,0)$, and 
$\frac{\sigma}{f_\pi^2}$ have changed several times, leading each time 
to updated force parameters $a'$, $b$, and $d$, according to Eqs.\ 
(\ref{E:TMa}) to (\ref{E:TMd}). We estimate here how these experimental 
uncertainties propagate into uncertainties of the triton binding energy 
$E_t$. 

For simplicity, we introduce the dimensionless variables
\begin{equation}
x=\frac{\sigma}{f_\pi^2} m_{\pi^+}\, , \quad
y=g^2 \, , \quad
z=\bar{F}^+(0,m_{\pi^+}^2) m_{\pi^+}\, , \quad
u=\bar{B}^-(0,0) m_{\pi^+}^2 \, ,
\end{equation}
\begin{equation}
\bar{\Lambda}=\frac{\Lambda}{\mu} \, , \quad
r=\frac{\mu}{m_{\pi^+}} \, , \quad
q=\frac{\mu}{m}\, , 
\end{equation}
as well as the dimensionless force parameters
\begin{equation}
\bar{a}=a' \mu \, , \quad
\bar{b}=b \mu^3 \, , \quad
\bar{d}=d \mu^3 \, .
\end{equation}
Using this notation, Eqs.\ (\ref{E:TMa}) through (\ref{E:TMd}) become
\begin{eqnarray}
\bar{a} & = & x r \left( 1-2r^2-2\frac{\bar{\Lambda}^2-1}{\bar{\Lambda}^4}
\right) + {1\over 2} q^3 y \, ,\label{E:TMa2}\\
\bar{b} & = & -2r^3(x-z) \, ,\label{E:TMb2}\\
\bar{d} & = & -{1\over 4} q^3 y -{1\over 2} r^2 q u \, .\label{E:TMd2}
\end{eqnarray}

Adding (independent) errors in quadrature, and taking into account the
relations (\ref{E:TMa2}) through (\ref{E:TMd2}), we arrive at the following
expression for the square of the uncertainty in $E_t$ due to the uncertainties
$\Delta x$, $\Delta y$, $\Delta z$, and $\Delta u$ in $x$, $y$, $z$, and $u$:

\begin{eqnarray}
\lefteqn{(\Delta E_t)^2 = 
\left[r\left(1-2r^2-2 \frac{\bar{\Lambda}^2-1}{\bar{\Lambda}^4}\right) 
\frac{\partial E_t}{\partial\bar{a}}
-2r^3 \frac{\partial E_t}{\partial\bar{b}} \right]^2 (\Delta x)^2 +}  
\nonumber\\
& & 
\left[{1\over 2} q^3 \left( \frac{\partial E_t}{\partial\bar{a}} 
-{1\over 2}\frac{\partial E_t}{\partial\bar{d}} \right) \right]^2
(\Delta y)^2
+\left[2r^3\frac{\partial E_t}{\partial\bar{b}}\right]^2 (\Delta z)^2
+\left[{1\over 2}r^2 q \frac{\partial E_t}{\partial\bar{d}} \right]^2 
(\Delta u)^2 \, . \label{E:error}
\end{eqnarray}
Note that it is simpler to vary the three constants $\bar{a}$,  
$\bar{b}$, and $\bar{d}$ instead of the four experimental values $x$, 
$y$, $z$, and $u$, but one has to keep in mind that the former are not 
independent from each other.

The partial derivatives $\partial E_t/\partial \bar{a}$, $\partial 
E_t/\partial \bar{b}$, and $\partial E_t/\partial \bar{d}$, are 
calculated numerically for the standard set of parameters of Table 
\ref{T:TMconst} at  $\bar{\Lambda}=5.8$. As estimates of the 
experimental uncertainties in $x$ and $z$ we use the values given in 
Ref.\ \cite{CoonHan}, $\Delta x = 0.25$ and $\Delta z = 0.05$. As a 
reasonable estimate for $\Delta y$ we choose the difference between the 
current value $y=172.1$ and the one used in the original TM force, 
$y=179.7$, yielding $\Delta y = 7.6$. The value $u=8.1$ is given in 
\cite{CoonHan} without indication of the error. Assuming that the 
specified digits are indeed significant, we set $\Delta u=0.1$. 

\begin{table}[bt]
\begin{center}
\begin{ruledtabular}
\begin{tabular}{lcccc}
  & Reid & Paris & Nijmegen 93 & Bonn B \\
\hline
$\partial E_t/\partial \bar{a}$ & \hspace{5mm}0.080\hspace{5mm}  & \hspace{5mm} 0.090
\hspace{5mm}    &\hspace{5mm} 0.095\hspace{5mm} & \hspace{5mm}0.095\hspace{5mm} \\
$\partial E_t/\partial \bar{b}$ & 0.725 & 0.745 & 0.845 & 0.750 \\
$\partial E_t/\partial \bar{d}$ & 0.545 & 0.470 & 0.540 & 0.505 \\
$\Delta E_t $ & 0.377 & 0.389 & 0.440 & 0.393 \\
$\Delta E_t^\mathrm{app}$ & 0.362 & 0.372 & 0.423 & 0.375\\
\end{tabular}
\end{ruledtabular}
\caption{Numerical results for the partial derivatives of the triton 
binding energy with respect to the TM $\pi-\pi$ force parameters and 
resulting uncertainties according to Eqs.\ (\ref{E:error}) and 
(\ref{E:error-ap}).} 
\label{T:errors} 
\end{center}
\end{table}

The numerical results for the partial derivatives of $E_t$ with respect 
to $\bar{a}$, $\bar{b}$, and $\bar{d}$, and the corresponding values of 
$\Delta E_t$ are shown in Table \ref{T:errors} for the TM $\pi-\pi$ 
force in combination with various $NN$ potentials. In each case, 
$\Delta E_t$  is roughly 0.4 MeV. This is a significant value which 
clearly shows that the dependence on $\Lambda_{\pi NN}$ is not the only 
source of uncertainty in predictions of $E_t$. 

A closer inspection of (\ref{E:error}) reveals that $\Delta E_t$ is 
almost completely dominated by $\Delta x$, the uncertainty in the 
pion-nucleon sigma term. One can easily derive the approximate form 
\begin{equation}
\Delta E_t^\mathrm{app} \approx 2 \frac{\partial E_t}{\partial\bar{b}} 
\Delta x \, , \label{E:error-ap}
\end{equation}
the results of which are also shown in Table \ref{T:errors} and come 
very close to the full result, confirming that the uncertainties in the 
other parameters are secondary as long as the  pion-nucleon sigma term 
is not determined with significantly better accuracy. 

Since (\ref{E:error-ap}) is a lower limit of (\ref{E:error}) one also 
has to conclude that currently {\em predictions} of the triton binding 
energy employing the TM force cannot be made with a better accuracy 
than plus or minus 0.4 MeV. Clearly, adding other contributions to the 
$3N$ force will further increase this uncertainty. 

\subsection{$\pi-\rho$ exchange force}

In the $NN$ potential, $\rho$-meson exchange provides important 
contributions to the tensor and spin-orbit components.  Its role is 
enhanced by the large anomalous $\rho NN$ coupling $\kappa_\rho$, 
connected (via VMD) to the anomalous isovector magnetic moment of the 
nucleon. The $\pi-\rho$ exchange was therefore always considered  the 
next most important part of the $3N$ force  
\cite{Ellis,PenaCoon,Stadler} after $\pi-\pi$ exchange. In this paper 
we will take another look at its contributions stemming from the 
nucleon Born diagrams. The TM $\pi-\rho$ exchange force also includes 
processes with intermediate $\Delta$ resonance excitations, which 
numerically turn out to be equally important. In our calculations the 
$\pi-\rho$ potentials with intermediate $\Delta$ resonance are included 
in the form (and with the corresponding parameters) specified in Ref.\ 
\cite{Stadler}.  

The nucleon Born contributions to the $\pi-\rho$ $3N$ force are derived 
in Appendix \ref{A:Born}. They depend significantly on the form of the 
$\pi NN$ coupling. For the PV $\pi NN$ coupling one gets 
\begin{eqnarray}
W^{\text{PV+}}_{\pi\rho T} (1) &=& 
 -(2 \pi )^3 {g^2 g_\rho^2\over 4 m^3} 
 \Bigrb[ (\bm{\tau}_2 \cdot \bm{\tau}_3) \bm{\sigma}_1 \cdot {\bf q}_3
 - (i \bm{\tau}_1 \cdot \bm{\tau}_2 \times \bm{\tau}_3)
 \bm{\sigma}_1 \cdot {\bf Q}_1 \Biglb] \nonumber\\
&& \quad  \times \quad \bm{\sigma}_2 \cdot {\bf q}_2 \,
   \tilde{\Delta}_\pi({\bf q}_2)  \tilde{\Delta}_\rho({\bf q}_3) 
 + 2 \leftrightarrow 3 \, , 
\label{pirhoPVT}\\
W^{\text{PV+}}_{\pi\rho S} (1) &=&  0 \, , \label{pirhoPVS}
\end{eqnarray}
whereas for the PS $\pi NN$ coupling
\begin{eqnarray}
W^{\text{PS-}}_{\pi\rho T} (1) &=& 
 -(2 \pi )^3 {g^2 g_\rho^2 (1+ \kappa_\rho)\over 4 m^3} 
 (\bm{\tau}_2 \cdot \bm{\tau}_3) 
  \bm{\sigma}_1 \cdot {\bf q}_3\, \bm{\sigma}_2 \cdot {\bf q}_2 
  \tilde{\Delta}_\pi({\bf q}_2)  \tilde{\Delta}_\rho({\bf q}_3) \, 
 + 2 \leftrightarrow 3 \, , 
\label{pirhoPST}\\
W^{\text{PS-}}_{\pi\rho S} (1) &=&  +(2 \pi )^3 {g^2 g_\rho^2\over 4 m^3}\,   
(i \bm{\tau}_1 \cdot \bm{\tau}_2 \times \bm{\tau}_3)
 \Bigrb[ \bm{\sigma}_1 \cdot {\bf Q}_3 +
 (1+\kappa_\rho) 
 i \bm{\sigma}_1 \times \bm{\sigma}_3 \cdot {\bf q}_3 \Biglb]
 \nonumber\\
&&  \quad  \times \quad \bm{\sigma}_2 \cdot {\bf q}_2 \,
   \tilde{\Delta}_\pi({\bf q}_2)  \tilde{\Delta}_\rho({\bf q}_3) 
 + 2 \leftrightarrow 3 \, , 
\label{pirhoPSS} 
\end{eqnarray}
where again ${\bf q}_i= {\bf p}_i^\prime- {\bf p}_i$, ${\bf Q}_i= {\bf 
p}_i^\prime+ {\bf p}_i$. The subscripts $S$ and $T$ refer to the 
exchange of the space and time components of the $\rho$ field, 
respectively. For brevity, we will call the corresponding terms 
``spacelike'' and ``timelike''. The superscripts PS and PV correspond 
to the type of $\pi NN$ coupling, the superscripts ``+'' stand for the 
contribution of the positive energy nucleon Born diagrams as derived in 
Appendix \ref{A:Born}, superscripts ``--'' denote the ``true'' pair (or 
Z-diagram) terms. 

In the numerical calculations we keep the $\pi-\rho$ force parameters of
\cite{Ellis,Stadler}, with the exception of $g^2$ which is updated to the
value of TM'(99) (see Table \ref{T:3NFconstants}).

\begin{table}[bt]
\begin{center}
\begin{ruledtabular}
\begin{tabular}{llccc}
Meson & mass (MeV)  &$\frac{g^2}{4 \pi}$ &$\frac{g^{*\, 2}}{4\pi}$ & 
cut-off\\ \hline 
 $\pi$   & 138   & 13.6953 &  2.1664 & $\Lambda_{\pi NN}=\Lambda_{\pi NN^*}$=5.8$m_\pi$ \\   
 $\rho$  & 768.3 & 0.81    &  & $\Lambda_{\rho NN}$ (Dirac)=12$m_\pi$
\\             
  & & & & $\Lambda_{\rho_NN}$(Pauli)=7.4$m_\pi$ \\
 $\sigma$ & 584 & 10.3251 & 0.6 & $\Lambda_{\sigma NN}=\Lambda_{\sigma NN^*}$=1995 MeV \\
 $\omega$& 782.6& 24.5&1.4237& $\Lambda_{\omega NN}=\Lambda_{\omega NN^*}$=1850 MeV \\
\end{tabular}
\end{ruledtabular}
\end{center}
\caption{Masses, couplings, and cut-off parameters that appear in the 
$3N$ forces. For the $\rho$ and the $\omega$ mesons there are also the 
tensor/vector coupling ratios $\kappa_\rho=6.6$ and 
$\kappa_\omega=0.0$, respectively. The parameters of the $\pi$ and 
$\rho$ mesons  are taken from the TM'(99) force and from Ref.\ 
\cite{Stadler}, the parameters of the $\omega$ meson from the Bonn B 
potential. The $\sigma$ meson parameters are determined in Appendix 
\ref{A:sigma}.} \label{T:3NFconstants} 
\end{table}

The terms proportional to ${\bf Q}_i$ are non-local and have been 
omitted so far from $3N$ potentials, mainly because of the difficulties 
associated with performing calculations with non-local interactions in 
coordinate space. Working in momentum space, we are in principle not 
hampered by non-localities. However, in this paper we focus on the 
relation between traditional meson-theoretical $3N$ forces and the $3N$
forces derived from ChPT where non-local terms are discarded. Hence it 
would not be useful to keep those terms in our calculations, and we 
neglect them as well. 

The above results illustrate one important point: One cannot rely on 
the nucleon Born terms alone to yield the most important contribution 
to the $3N$ force. In particular, it would be incorrect to claim that 
since the PV $\pi NN$ coupling is ``more consistent'' with the 
important chiral symmetry requirement, one should use the above 
potentials $W^{\text{PV+}}_{\pi\rho T}$ and $W^{\text{PV+}}_{\pi\rho 
S}$. Indeed,  the construction of the $\pi-\rho$ $3N$ force by the TM 
group \cite{Ellis} via the extension of CA to include vector meson 
dominance, has shown that its most important part - the so called 
Kroll-Ruderman (KR) term - follows (in the theory with PV coupling) not 
from the nucleon Born term, but from an additional contact term 
required by gauge invariance of the pion photoproduction amplitude. 
This is, of course, very similar to the situation for the leading order 
isovector meson exchange currents: there the nucleon Born term with PS 
coupling gives the most important contribution (which is supplemented 
by the pion-in-flight diagram), whereas the Born term for PV coupling 
does not contribute and gauge invariance requires the presence of the 
contact interaction. The KR contact term happens to be opposite in sign 
to the chiral  contact term, arising from the chiral rotation from the 
PV to PS coupling and introduced in Appendix \ref{A:Born}. Thus, it appears 
reasonable to adopt for the $\pi-\rho S$ exchange potential just that 
KR term, i.e., to take 
\begin{equation}
W_{\pi\rho S}^{KR}= -
     W^{\text{cont}}_{\pi\rho S}= W^{\text{PS-}}_{\pi\rho S}
\end{equation}
But then it would seem natural to take for the timelike exchange the
corresponding $W_{\pi\rho T}=W^{\text{PS-}}_{\pi\rho T}$, which is 
enhanced by the large factor $1+\kappa_\rho$ compared to 
$W^{\text{PV+}}_{\pi\rho T}$. Note, that in \cite{CPR} (in which the 
Born contributions to the $\pi-\rho S$ where first derived) an 
expression identical to $W^{\text{PV+}}_{\pi\rho T}$ is listed for the 
timelike part, while $W^{\text{PS-}}_{\pi\rho S}$ (the KR term) is 
given for the spacelike part \footnote{There are, unfortunately, two 
misprints in Eq.\ (2.8) of Ref.\ \cite{CPR}: a factor of 
$(1+\kappa_\rho)$ is missing in the second term and the first term has 
an extra factor 2.}. 

We do not intend to imply by the discussion above that the PS $\pi NN$
coupling is to be preferred in this context. In fact, trying to make 
any preference for the Lagrangian involving heavy mesons is meaningless 
(unless one attempts to extend the global chiral symmetry to a local 
one and builds a model which provides all relevant contact terms 
\cite{Smejkal}). We only want to point out that more simplistic 
approaches (like the one of this paper and that of Ref.\ \cite{CPR}: 
taking only the Born terms with a hope that they contain the most 
important effects) do not give an unambiguous prediction. Besides, the 
$\rho-T$ contribution can be comparable to others when it is enhanced  
by the large factor $1+\kappa_\rho$, as it happens for PS coupling. 

\subsection{$\pi-\sigma$ and $\pi-\omega$ exchange force}

Since $\sigma$ and $\omega$ meson exchanges play an important role in 
OBE models of the $NN$ interaction, they contribute naturally also to 
the $3N$ force. Short-range $\pi-\sigma$ and $\pi-\omega$  $3N$ 
potentials were introduced in Ref.\ \cite{CPR}. They are derived from 
diagrams with intermediate positive-energy nucleons with PV $\pi NN$ 
coupling.  

Unlike in the $\pi-\pi$ force, the Born terms in $\pi-\sigma$ and 
$\pi-\omega$ potentials (Fig.\ \ref{F:short-longandlong-long}a) are  
rather large. As the $\pi-\rho$ potentials from the previous section, 
they depend on the type of $\pi NN$ coupling. We also include 
additional $\sigma$ and  $\omega$ exchange contributions  generated by 
excitations of the intermediate nucleon to the Roper resonance 
\cite{CPR} (Fig.\ \ref{F:short-longandlong-long}b). 

The $\pi-\sigma$ and $\pi-\omega$ potentials corresponding to the  
nucleon Born diagrams follow immediately from the expressions developed 
in Appendix \ref{A:Born}. For $\pi-\sigma$ exchange we get 
\begin{eqnarray} 
W^{\text{PV+}}_{\pi\sigma} (1) &=&  +(2 \pi )^3
{g^2g_\sigma^2\over 4 m^3}  (\bm{\tau}_1 \cdot \bm{\tau}_2)\, \bm{\sigma}_1 \cdot {\bf
q}_3 \, \bm{\sigma}_2 \cdot {\bf q}_2\, \tilde{\Delta}_\pi({\bf q}_2) 
\tilde{\Delta}_\sigma({\bf q}_3) \,  + 2 \leftrightarrow 3 \, , \,  
\label{pisigmaPV}\\ 
W^{\text{PS-}}_{\pi\sigma} (1) &=&  -(2 \pi )^3 {g^2g_\sigma^2\over 4 m^3}  (\bm{\tau}_1
\cdot \bm{\tau}_2)\, \bm{\sigma}_1 \cdot {\bf q}_2 \, \bm{\sigma}_2 \cdot {\bf q}_2\,
\tilde{\Delta}_\pi({\bf q}_2)  \tilde{\Delta}_\sigma({\bf q}_3) \,  + 2 \leftrightarrow 3
\, ,      \label{pisigmaPS} 
\end{eqnarray} 
and for $\pi-\omega$ exchange 
\begin{eqnarray} 
W^{\text{PV+}}_{\pi\omega} (1) &=&  -(2 \pi )^3 {g^2g_\omega^2\over 4
m^3}  (\bm{\tau}_1 \cdot \bm{\tau}_2)\, \bm{\sigma}_1 \cdot {\bf q}_3 \, \bm{\sigma}_2
\cdot {\bf q}_2\, \tilde{\Delta}_\pi({\bf q}_2)  \tilde{\Delta}_\omega({\bf q}_3) \,  + 2
\leftrightarrow 3 \, , \,  
\label{piomegaPV}\\ 
W^{\text{PS-}}_{\pi\omega} (1) &=&  -(2
\pi )^3 {g^2g_\omega^2 (1+\kappa_\omega)\over 4 m^3}  (\bm{\tau}_1 \cdot \bm{\tau}_2)\,
\bm{\sigma}_1 \cdot {\bf q}_3 \, \bm{\sigma}_2 \cdot {\bf q}_2\, \tilde{\Delta}_\pi({\bf
q}_2)  \tilde{\Delta}_\omega({\bf q}_3) \,  + 2 \leftrightarrow 3 \, .     
\label{piomegaPS} 
\end{eqnarray} 
In Ref.\ \cite{CPR}, only the potentials for PV coupling are given.  
Since $\kappa_\omega$ is very small, the results for $\pi-\omega$  do 
not depend much on the type of the $\pi NN$ coupling. However, the  
$\pi-\sigma$ potentials do differ. For  PV coupling, the $\pi-\sigma$ 
and $\pi-\omega$  potentials $W^{\text{PV+}}_{\pi\sigma} (1)$ and   
$W^{\text{PV+}}_{\pi\omega} (1)$ have identical structure but opposite 
sign. Therefore, as in the case of the $NN$ interaction, a strong 
cancellation between the $\pi-\sigma$ and $\pi-\omega$ $3N$ interactions 
occurs. For small momenta, the sum of these potentials is proportional 
to  
\begin{equation}
\frac{g^2_\sigma}{m_\sigma^2}-\frac{g^2_\omega}{m_\omega^2} \,  ,
\label{sigomratio}
\end{equation} 
just as for the corresponding $NN$ potentials. For the potentials with 
PS coupling, Eqs.\ (\ref{pisigmaPS},\ref{piomegaPS}), such a 
cancellation does not take place. However, they would cancel if the 
potential (\ref{piomegaPS}) did not change much  when the momentum 
${\bf q}_3$ is replaced  by the $-{\bf q}_2$. As discussed later, this 
is exactly the momentum replacement needed to extract the low-energy 
constants (LECs) of ChPT. Therefore we calculate the $\pi-\sigma$ 
potential in  both forms. The extent of the cancellations between  
$\sigma$ and  $\omega$ exchanges depends also on the numerical values 
of the masses  and coupling constants of the $\sigma$ and $\omega$ 
mesons. 

While we are trying to parameterize the $3N$ forces as consistent as 
possible with the $NN$ potentials they are combined with, we face a 
problem with the $\sigma$ meson. The only $NN$ potential with explicit 
$\sigma$ exchange we are using is Bonn B. Its $\sigma$ meson, however, 
is not a pure scalar-isoscalar particle, but has also a 
scalar-isovector component. This is reflected in the fact that the 
$\sigma$ mass, the $\sigma NN$ coupling constant and cut-off mass is 
different in $NN$ isospin 0 and 1 channels. 

In order to get a true OBE representation of $\sigma$-exchange, we keep 
for the $3N$ potential only the part of the Bonn $\sigma$ meson which 
corresponds to pure isoscalar exchange. The parameters of our $\sigma$ 
are then related to the Bonn $\sigma$ such that a simple OBE form, 
consistent with the way it is implemented in the $3N$ forces, is 
obtained. Details of this procedure can be found in Appendix 
\ref{A:sigma}.

The forces originated from excitations of the Roper resonance read  
\begin{eqnarray} 
W_{\pi\sigma}^*(1)&=& - (2 \pi )^3 
{g g^* g_\sigma g_\sigma^*\over 2(m^*-m)m^2}\,
  ( \bm{\tau}_1 \cdot \bm{\tau}_2 )\, 
 \bm{\sigma}_1 \cdot {\bf q}_2 \, \bm{\sigma}_2 \cdot {\bf q}_2\,
  \tilde{\Delta}_\pi({\bf q}_2)  \tilde{\Delta}_\sigma({\bf q}_3) \, 
 + 2 \leftrightarrow 3 \, ,
\label{pisigmar} \\
W_{\pi\omega}^*(1)&=& + (2 \pi )^3  
{g g^* g_\omega g_\omega^*\over 2(m^*-m)m^2}\,
 ( \bm{\tau}_1 \cdot \bm{\tau}_2 )\,
 \bm{\sigma}_1 \cdot {\bf q}_2 \, \bm{\sigma}_2 \cdot {\bf q}_2\,
  \tilde{\Delta}_\pi({\bf q}_2)  \tilde{\Delta}_\omega({\bf q}_3) \, 
 + 2 \leftrightarrow 3 \, .
\label{piomegar}
\end{eqnarray}
Here $m^*$ is the mass of the $N^*(1440)$ resonance. We would like to 
stress that the form of these potentials depends on the structure of 
the $NN^*\sigma$  and $NN^*\omega$ vertices. We adopted the very simple 
choice used in Ref.\ \cite{CPR}. The  form of these effective vertices 
follows from the quantum numbers of  the baryon state $N^*$ and of the 
$\sigma$ and $\omega$ mesons, neglecting any possible dependence on the 
substructure of these particles.  Even for the simple choice of the 
$N^*$ vertices of \cite{CPR}, theoretical predictions of the 
corresponding coupling constants (and  their ratios) from quark models 
are hampered by the uncertain quark content of $N^*$, whereas their 
extraction from experimental data can be done only in an indirect and 
model-dependent way. 

The strength of the $\pi NN^*$ coupling is calculated from the partial 
decay width of $N^* \rightarrow N+ \pi$. The recent determination 
\cite{Soyeur} based on the value $\Gamma (N^* \rightarrow N\pi )= 228$ 
MeV gives\footnote{Equation\ (3.3.) of Ref.\ \cite{CPR} is incorrect 
and the corresponding value of $\frac{f_{\pi NN^*}^2}{4\pi}$ is 
overestimated by a factor of 3.} 
\begin{equation}
 \frac{f_{\pi NN^*}^2}{4\pi}= 0.0117 \, , \quad \quad
 g^*= \frac{2m}{m_\pi}\, f_{\pi NN^*}= 5.22 \, .
\label{gstar}
\end{equation}

The couplings for the $\sigma NN^*$ and $\omega NN^*$ vertices are more 
difficult to pin down. The  coupling constant of $\sigma NN^*$, extracted 
from the partial decay width of $N^* \rightarrow N+ (\pi 
\pi)^{I=0}_{S-wave}$, depends critically on the assumed mass and  width 
of the $\sigma$ resonance \cite{Soyeur}. The ``$\sigma$ meson'' (with 
zero width) employed in the parameterization of the OBE potentials 
simulates not only $(\pi \pi)^{I=0}_{S-wave}$, but also other 
scalar-isoscalar exchanges. Such a phenomenological $\sigma$ exchange was 
used in a recent analysis \cite{Oset} of the inelastic scattering $\alpha 
+p \rightarrow \alpha + X$, from where the effective coupling constant 
of $\sigma NN^*$  was extracted as 
\begin{equation}
  \frac{g_\sigma^{*\, 2}}{4\pi}= 1.33 \, ,
\label{gsigstar}
\end{equation}
with $m_\sigma= 550$. This is much larger than typical values 
obtained from the Roper resonance decay \cite{Soyeur}.  

On the other hand, Ref.\ \cite{CPR} extracts $\frac{g_\sigma^{*\, 
2}}{4\pi}= 0.1$, which is even smaller than all values of Ref.\ 
\cite{Soyeur}. Given this wide spread of coupling constants in the 
literature, we adopt the intermediate $\frac{g_\sigma^{*\, 2}}{4\pi}= 
0.6$ as our standard value, but calculate also the two extreme cases in 
order to observe the sensitivity of the results with respect to the 
choice of this parameter. 

For $\omega NN^*$  we follow  Ref.\ \cite{CPR} and determine 
$g_\omega^*$ from the ratio 
\begin{equation}
  g_\omega^*= g_\omega\, \frac{g_\sigma^*}{g_\sigma} \, ,
\label{gomstar} 
\end{equation}
which follows from a naive constituent quark model.
The sum of the $3N$ potential with intermediate Roper resonance and 
$\pi-\sigma$ and $\pi-\omega$ exchanges is then roughly proportional to
\begin{equation}
- \frac{g_\sigma^*}{g_\sigma}\, \left(
\frac{g^2_\sigma}{m_\sigma^2}-\frac{g^2_\omega}{m_\omega^2} \right) \, .
\label{sigomstratio} 
\end{equation}
Therefore we can expect the same amount of cancellation between $\sigma$ 
and $\omega$ terms in $3N$ forces with Roper excitations as in $NN$ 
potential and $3N$ Born contributions. However, since the simple 
scaling rule of coupling constants is theoretically not very well 
founded, we also use sets of parameters that do not satisfy 
(\ref{gomstar}).

\subsection{Numerical results}

We calculated the triton binding energies and $3N$ wave functions for 
Hamiltonians containing both $NN$ and $3N$ potentials by solving the 
non-relativistic $3N$ Faddeev equations in momentum space exactly. 

The $3N$ potentials are expressed in terms of Jacobi momenta in the 
{\em cm} frame of the $3N$ system, and decomposed into partial waves in 
a basis of $jj$ coupling states. Because of the complexity of the 
resulting numerical problem, we restrict ourselves to those partial 
waves in which the $NN$ pair total angular momentum does not exceed 2, 
which corresponds to 18 different three-body channels. Details of the 
formalism and the numerical methods are described in Ref. 
\cite{Stadler91}. 

The main purpose of these calculations is to compare the effects of the 
various contributions to the $3N$ force which were described in the 
previous sections on the triton binding energy. In order to study also 
the dependence of our results on the $NN$ interaction included in the 
Hamiltonian, we used a number of different $NN$ potentials, namely Reid 
\cite{Reid}, Paris \cite{Paris}, Nijmegen 93 \cite{Nijmegen}, and 
Bonn B \cite{BonnB}. 

\begin{table}[bt]
\begin{center}
\begin{ruledtabular}
\begin{tabular}{ldddddddd}
 & \multicolumn{2}{c}{Reid} & \multicolumn{2}{c}{Paris} & \multicolumn{2}{c}{Nijmegen 93} & 
 \multicolumn{2}{c}{Bonn B} \\
3NF & \multicolumn{1}{c}{$E_t$} & \multicolumn{1}{c}{$\Delta E_t$} & \multicolumn{1}{c}{$E_t$} & 
\multicolumn{1}{c}{$\Delta E_t$} & \multicolumn{1}{c}{$E_t$} & \multicolumn{1}{c}{$\Delta E_t$} & 
\multicolumn{1}{c}{$E_t$} & \multicolumn{1}{c}{$\Delta E_t$} \\
\hline
  no 3NF & -7.230 &     & -7.383  &     &  -7.756  &        &  -8.100 &     \\
  $+\pi\pi(a')$ & -7.279 & -0.049 & -7.439  &  -0.056 &  -7.811  & -0.055 &  -8.159 & -0.059  \\
  $+\pi\pi(b)$ & -8.739 & -1.460 & -8.939  &  -1.500 &  -9.471  & -1.660 &  -9.624 & -1.465  \\
  $+\pi\pi(d)$ & -9.100 & -0.361 & -9.220  &  -0.281 &  -9.782  & -0.311 &  -9.847 & -0.223  \\
 $+\pi\rho({\mathrm KR})$ & -9.017 & 0.083  & -9.118  &  0.102  &  -9.635  & 0.147  &  -9.672 & 0.175   \\
 $+\pi\rho(\Delta^+)$ & -8.849 & 0.168  & -8.961  &  0.157  &  -9.464  & 0.171  &  -9.506 & 0.166   \\
 $+\pi\rho(\Delta^-)$ & -8.747 & 0.102  & -8.821  &  0.140  &  -9.285  & 0.179  &  -9.325 & 0.181   \\
 $+\pi\rho (T)$ & -8.772 & -0.025 & -8.850  &  -0.029 &  -9.316  & -0.031 &  -9.352 & -0.027  \\
 $+\pi\sigma (Z)$ & -8.273 & 0.499  & -8.213  &  0.637  &  -8.663  & 0.653  &  -8.658 & 0.694   \\
 $+\pi\sigma (N^*)$ & -8.711 & -0.438 & -8.610  &  -0.397 &  -9.145  & -0.482 &  -9.055 & -0.397  \\
 $+\pi\omega (Z)$ & -9.213 & -0.502 & -9.380  &  -0.770 &  -9.977  & -0.832 &  -9.956 & -0.901  \\
 $+\pi\omega (N^*)$ & -8.735 & 0.478  & -8.898  &  0.482  &  -9.370  & 0.607  &  -9.524 & 0.432   \\
\end{tabular}
\end{ruledtabular}
\caption{Triton binding energies and their differences (in MeV) 
calculated for various model Hamiltonians with different $NN$ 
potentials and contributions to the $3N$ force added consecutively. All 
$\pi NN$ vertices in the $3N$ forces of this table are calculated in PV 
coupling. The columns labeled $E_t$ show the triton binding energies, 
while the ones labeled $\Delta E_t$ indicate the differences between 
the binding energies of consecutive rows, indicating the effect of the 
corresponding $3N$ force component.} 
\label{T:EPV} 
\end{center}
\end{table}

In Tables \ref{T:EPV} and \ref{T:EPS} we show the triton binding 
energies for various Hamiltonians where the $3N$ forces of the previous 
sections are added successively. The binding energy differences depend 
somewhat on the order in which the potentials are added, therefore we 
later also show their expectation values. 

\begin{table}[hbt]
\begin{center}
\begin{ruledtabular}
\begin{tabular}{ldddddddd}
 & \multicolumn{2}{c}{Reid} & \multicolumn{2}{c}{Paris} & \multicolumn{2}{c}{Nijmegen 93} & 
 \multicolumn{2}{c}{Bonn B} \\
3NF & \multicolumn{1}{c}{$E_t$} & \multicolumn{1}{c}{$\Delta E_t$} & \multicolumn{1}{c}{$E_t$} & 
\multicolumn{1}{c}{$\Delta E_t$} & \multicolumn{1}{c}{$E_t$} & \multicolumn{1}{c}{$\Delta E_t$} & 
\multicolumn{1}{c}{$E_t$} & \multicolumn{1}{c}{$\Delta E_t$} \\
\hline
 $+\pi\rho (T)$     & -8.859  & -0.112 & -8.953   & -0.132 & -9.425   & -0.140 & -9.452  & -0.127  \\
 $+\pi\sigma (Z)$   & -10.492 & -1.633 & -10.544  & -1.591 & -11.431  & -2.006 & -10.879\footnote{In
 addition to the specified result, another unphysical, deeply bound solution was obtained in this case.} & -1.427  \\
 $+\pi\sigma (N^*)$ & -11.219 & -0.727 & -11.264  & -0.720 & -12.215\footnotemark[1] & -0.784 & 
 -11.600\footnotemark[1] & -0.721  \\
 $+\pi\omega (Z)$   & -13.680 & -2.461 & -15.367  & -4.103 & -20.952  & -8.737 & 
 -15.811\footnotemark[1] & -4.211  \\
 $+\pi\omega (N^*)$ & -12.674 & 1.006  & -14.134  & 1.233  & -16.869  & 4.083  & -18.345 & -2.534  \\
\end{tabular}
\end{ruledtabular}
\caption{Triton binding energies and their differences (in MeV), as in 
Table \ref{T:EPV}, but with $\pi NN$ PS coupling in the $3N$ forces. 
The binding energy differences in the first row are calculated with 
respect to the corresponding entries in the row labeled 
$+\pi\rho(\Delta^-)$ of Table \ref{T:EPV}. Note that only for $\pi\rho 
(T)$ and $\pi\sigma (Z)$ the $3N$ potentials using PS and PV coupling 
actually differ. The effects of the other, unchanged $3N$ potentials on 
the binding energy are amplified compared to the case of PV coupling, 
since the $3N$ wave functions are significantly altered by the $\pi\rho 
(T)$ and $\pi\sigma (Z)$ PS potentials.} 
\label{T:EPS} 
\end{center}
\end{table}

\begin{figure}
\begin{tabular}{c}
\epsfig{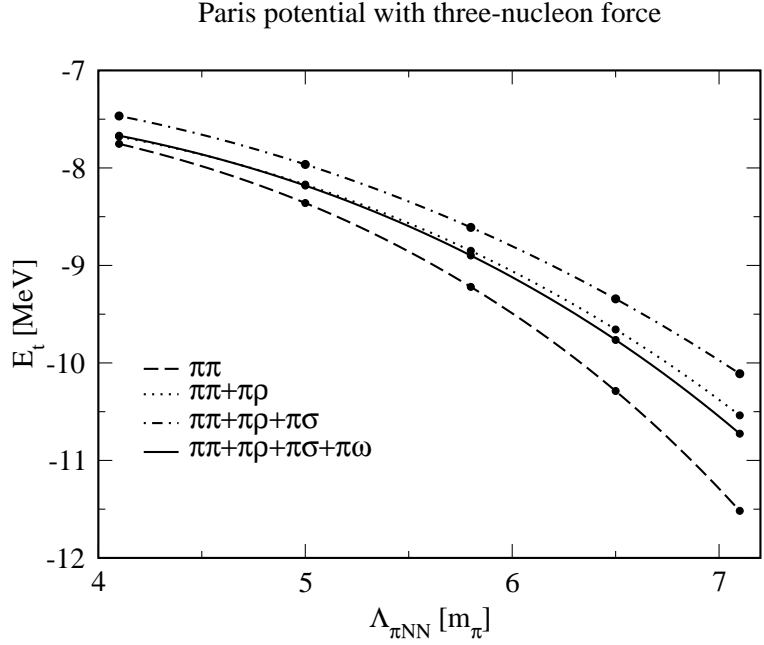} \\ 
\\
\epsfig{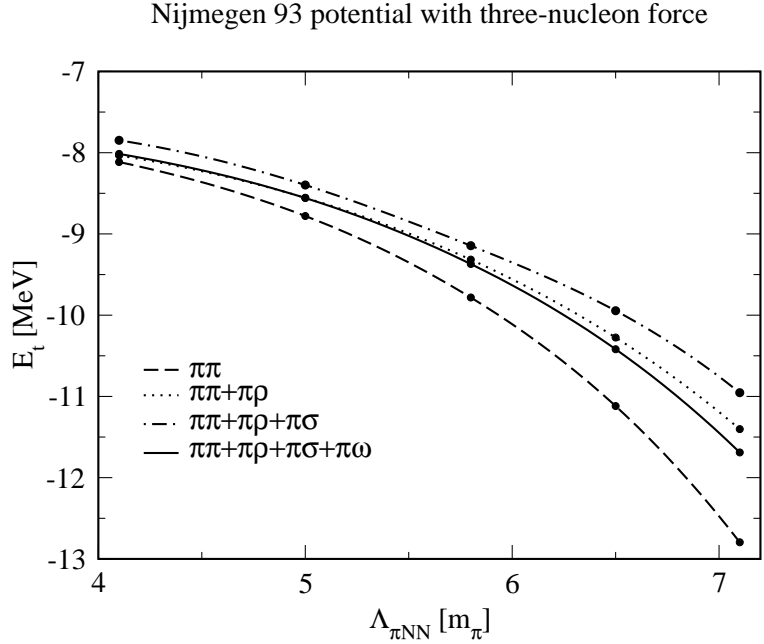} 
\end{tabular}
\caption{Dependence of the triton binding energy on the $\pi NN$ 
cut-off parameter, calculated for Hamiltonians with the Paris and the 
Nijmegen 93 $NN$ potentials and various contributions to the $3N$ 
force, which are successively added in the following way: $\pi-\pi$ 
exchange (dashed line), plus $\pi-\rho$ (dotted line), plus 
$\pi-\sigma$ (dashed-dotted line),  plus $\pi-\omega$ (solid line). The 
actual calculations are indicated by the full circles, the lines are 
drawn to guide the eye.} \label{F:figurecut} 
\end{figure}

First we make a few comments to the results with the $\pi-\pi$ $3N$ potential 
only. In our calculations \cite{Stadler} with the old version of the TM 
force TM(93), which contains the $c$-term and uses somewhat different 
values of the constants $a, b$ and $d$, we obtained $E_t(\mathrm{Reid})= -8.904$ MeV and 
$E_t(\mathrm{Paris})= -9.060$ MeV. This means that transforming the $c$-term away 
and using the new version TM'(99), as in Table \ref{T:EPV}, brings an extra 
binding of about 200 keV. The first line of Table \ref{T:nonpert} shows that, even 
after removing the singular $c$-term, the TM $\pi-\pi$ force remains 
highly non-perturbative. Also the cut-off dependence of $E_t$ (see Fig. 
\ref{F:figurecut}) remains about the same (see Refs.\ 
\cite{CoonHan,Stadler}) as with the old version TM(93).

\begin{table}[tb]
\begin{ruledtabular}
\begin{tabular}{lddddd}
$3N$ potential & \multicolumn{1}{c}{$\Delta E_t$} & \multicolumn{1}{c}{no $3N$ force} & 
\multicolumn{1}{c}{with $\pi\pi$} &  \multicolumn{1}{c}{with $\pi\pi$+$\pi\rho$} &
\multicolumn{1}{c}{with $\pi\pi$+$\pi\rho$+$\pi\sigma$+$\pi\omega$}\\
\hline
$\pi\pi$    & -1.837  & -1.281 & -2.572 & -2.190 & -2.274 \\
$\pi\rho$   & 0.370   & 0.139  & 0.450  & 0.293 & 0.317  \\
$\pi\sigma$     & 0.240   &  0.275 & 0.748 &  0.605 &  0.881 \\
$\pi\omega$     & -0.288  & -0.228 & -0.758 & -0.594  & -0.968 \\
\end{tabular}
\end{ruledtabular}
\caption{Matrix elements of the $3N$ potentials of the first column 
calculated with different wave functions. For comparison, the column 
labeled $\Delta E_t$ shows the binding energy differences according to 
Table \ref{T:EPV}. All wave functions are calculated with the Paris 
potential and various $3N$ forces in the Hamiltonian. } 
\label{T:nonpert} 
\end{table}

Most of the new short-range potentials considered in this paper  
contribute {\em individually} as much as (or more than) the TM 
$\pi-\rho$ short-range  forces considered before. The only exception is 
the  $\rho-T$ term, also not considered before. It is the only 
attractive part of  the $\pi-\rho$ $3N$ force, but it is about 5  times 
smaller than the KR term. Only in PS coupling the $\rho-T$ term is 
enhanced by the factor $1+ \kappa_\rho$ and becomes comparable to the 
KR. 

The $\sigma$ and $\omega$ exchange contributions are rather large. The 
Born terms are repulsive in case of the $\sigma$ and attractive for the 
$\omega$, but when PV $\pi NN$  coupling is used they cancel in a 
similar way as the corresponding $NN$ potentials. For the PS coupling, 
the $\pi-\sigma$ Born contribution turns attractive and together the  
$\pi-\sigma$ and $\pi-\omega$ exchanges lead to strong overbinding, as 
Table \ref{T:EPS} demonstrates.  Nevertheless, it would be premature to 
rule out the PS coupling solely based on these results: we would like 
to remind that other non-Born contact terms have not been investigated 
so far and are likely to be as important as the Born term considered 
here.  

\begin{table}[hbt]
\begin{center}
\begin{ruledtabular}
\begin{tabular}{ldddddd}
 & \multicolumn{2}{c}{$\frac{g^*_\sigma}{4\pi}=0.1$} & 
 \multicolumn{2}{c}{$\frac{g^*_\sigma}{4\pi}=0.6$} & 
 \multicolumn{2}{c}{$\frac{g^*_\sigma}{4\pi}=1.33$} \\
3NF & 
\multicolumn{1}{c}{$E_t$} & \multicolumn{1}{c}{$\Delta E_t$} & 
\multicolumn{1}{c}{$E_t$} & \multicolumn{1}{c}{$\Delta E_t$} & 
\multicolumn{1}{c}{$E_t$} & \multicolumn{1}{c}{$\Delta E_t$} \\
\hline
$\pi\pi+\pi\rho$ & -8.850 & & -8.850 & & -8.850 \\
$+\pi\sigma (Z)$ &
 -8.213 & 0.637 & -8.213 & 0.637 & -8.213 & 0.637 \\
 $+\pi\sigma (N^*)$     & 
 -8.369 & -0.156 & -8.610 & -0.397 & -8.822  & -0.609  \\
 $+\pi\omega (Z)$   & 
 -9.073 & -0.704 & -9.380 & -0.770 & -9.652  & -0.830  \\
\hline
with  $g_\omega^*= g_\omega\, \frac{g_\sigma^*}{g_\sigma}$ \\ 
 $+\pi\omega (N^*)$ & 
 -8.883 &  0.190 & -8.898 &  0.482 & -8.910  &  0.742 \\
\hline
with  $\frac{g^*_\omega}{4\pi}=0.24$\\
$+\pi\omega (N^*)$ &
-8.882 & 0.191 & -9.174 & 0.206 & -9.431 & 0.221 \\
 \end{tabular}
\end{ruledtabular}
\caption{Triton binding energies and their differences (in MeV), as in Table \ref{T:EPV},
for the Paris $NN$ potential and $3N$ potentials with different coupling constants
$g_\sigma^*$ and $g_\omega^*$. The first two lines are taken from Table \ref{T:EPV} to
define reference values. The contribution of $\pi\sigma (N^*)$ increases with increasing
$g_\sigma^*$, and so does the contribution of $\pi\omega (N^*)$ in the fifth line for a
fixed ratio $g_\omega^*/g_\sigma^*$, keeping the binding energy almost constant. When
$g_\omega^*$ is kept fixed (last line), these cancellations no longer take place and the
binding energy increases.}  
\label{T:Roper} 
\end{center}
\end{table}

One example of such contributions are forces with an intermediate Roper 
resonance. We observe that individually they are also rather large. 
However, as we have already pointed out above, there are significant 
uncertainties in the coupling constants involving $N^*$, and the effect 
of those potentials can vary accordingly. Nevertheless, it is 
remarkable that even for our smallest value of $g^*_\sigma$ the 
$\pi-\sigma N^*$ term is comparable to the $\pi-\rho$ $\Delta$ forces, 
as can be seen by comparing Tables \ref{T:EPV} and \ref{T:Roper}. 
Although the effect of the $\pi-\sigma N^*$ force scales with 
$g^*_\sigma$, the overall contribution of $\pi-\sigma N^*$ and 
$\pi-\omega N^*$ forces does not change much, since we assume the 
constituent quark model ratio $g^*_\sigma/g^*_\omega$, which leads to a 
strong cancellation  between them. Removing this constraint, as for 
instance in the last line of Table \ref{T:Roper}, can have  an effect 
on the overall contribution of $\pi-\sigma$ and $\pi-\omega$ forces 
comparable to the transition from PV to PS coupling in Z-digrams: the 
cancellation between $\pi-\sigma N^*$ and $\pi-\omega N^*$ forces could 
become much weaker or disappear and very pronounced changes of $E_t$ 
can be expected. Having more reliable information on 
$g_\omega^*/g_\sigma^*$ is thus absolutely crucial for more definite 
conclusion on the importance of the short-range $3N$ forces considered 
in this paper.   

\begin{table}[tb]
\begin{ruledtabular}
\begin{tabular}{ldddd}
$3N$ potential & \multicolumn{1}{c}{Reid} &\multicolumn{1}{c}{Paris} & 
\multicolumn{1}{c}{Nijmegen 93} &  \multicolumn{1}{c}{Bonn B} \\ \hline 
$\pi\pi(a')$ &  -0.096  &    -0.113  &    -0.124  &    -0.130  \\ 
$\pi\pi(b)$  &  -1.822  &    -1.882  &    -2.086  &    -1.837  \\ 
$\pi\pi(d)$  &  -0.364  &    -0.279  &    -0.319  &    -0.256  \\ 
\hline 
 Total $\pi\pi$ & -2.282 & -2.274 & -2.528 & -2.223 \\
\hline $\pi\rho({\mathrm KR})$  &  0.055   &    0.063   &    0.097   &    
0.113   \\ $\pi\rho(\Delta^+)$  &  0.156   &    0.148   &    0.164   &    
0.173   \\ 
 $\pi\rho(\Delta^-)$ &  0.101   &    0.142   &    0.179   &    0.174   \\
$\pi\rho (T)$  &  -0.026  &    -0.036  &    -0.036  &    -0.037  \\ 
\hline 
 Total $\pi\rho$ & 0.287 & 0.317 & 0.403 & 0.424 \\
\hline $\pi\sigma (Z)$  &  0.959   &    1.359   &    1.632   &    2.108   
\\ $\pi\sigma (N^*)$  &  -0.497  &    -0.478  &    -0.581  &    -0.381  
\\ \hline 
 Total $\pi\sigma$ & 0.463 & 0.881 & 1.051 & 1.727 \\
\hline $\pi\omega (Z)$  &  -0.902  &    -1.422  &    -1.722  &    
-2.377  \\ $\pi\omega (N^*)$  &  0.456   &    0.453   &    0.561   &    
0.371   \\ \hline 
 Total $\pi\omega$ & -0.446 & -0.968 &  -1.162 & -2.005 \\
\hline Total $\pi\sigma+\pi\omega$ & 0.016 & -0.087 & -0.110 & -0.278 
\\ Total $3N$ potentials & -1.979 & -2.045 &  -2.236 & -2.077 \\ 
\end{tabular}
\end{ruledtabular}
\caption{Expectation values (in MeV) of the components of the $3N$ 
force in Hamiltonians with different $NN$ potentials, calculated with 
eigenfunctions of the full Hamiltonian containing all listed $3N$ 
potentials.} 
\label{T:expvalues} 
\end{table}

In Table \ref{T:expvalues} we show the expectation values calculated 
with the wavefunction obtained from the Hamiltonian with all $3N$ 
potentials in PV coupling. The expectation values of the $\pi-\pi$ 
potential and of the full $3N$ potential do not vary much, but the 
individual forces show a rather strong dependence on the $NN$ potential 
used. Our results (Tables \ref{T:nonpert} and \ref{T:expvalues}) for  
the total $\pi-\sigma$ and $\pi-\omega$ contributions differ from those 
of Ref.\  \cite{CPR}, where $\langle W(\pi-\sigma)\rangle= 1.003$ MeV 
and $\langle W(\pi-\omega)\rangle = -0.770$ MeV were  obtained with a 
wave function calculated with the Paris $NN$ potential only. This 
difference is a consequence of the different coupling constants and 
cut-off parameters used in Ref. \cite{CPR}. The results collected in 
Tables \ref{T:expvalues} and \ref{T:nonpert} show again that also the 
short-range $3N$ forces are non-perturbative: even for the small sum of 
all $\pi-\sigma$ and $\pi-\omega$ contributions we get, for example 
with the Paris potential, -0.048 MeV from Table \ref{T:EPV} and -0.087 
MeV from Table \ref{T:nonpert}.    

Finally, in Fig.\ \ref{F:figurecut} we represent the dependence of the 
binding energy for two different $NN$ models as a function of the $\pi NN$ 
cut-off parameter. The short-range forces do make $E_t$ less cut-off dependent, 
compared to the case when only the $\pi-\pi$ potential is included. But 
the overall effect of the $\pi-\sigma$ and $\pi-\omega$ potentials is rather 
small. We emphasize again that this sum depends crucially on the 
values of the coupling constants used in our calculations, in 
particular on the poorly determined ratio $g^*_\sigma/g^*_\omega$.  
 
%%%%%%%%%%%%%%%%%%%%%%%%%%%%%%%%%%%%%%%%%%%%%%%%%%%%%%%%%%%%%%%%%%%%%%%
%%%%%%%%%%%%%%%%%%%%%%%%%%%%%%%%%%%%%%%%%%%%%%%%%%%%%%%%%%%%%%%%%%%%%%%
\section{Determination of the low energy constants}
\subsection{Effective couplings}

To connect to the $3N$ force following from ChPT, let us now consider
``the pointlike limit'', i.e., shrinking the propagators of heavy 
mesons to a point by taking  $m_B^2 \tilde{\Delta}_B({\bf q}_3) 
\rightarrow 1$ ($B= \rho, \omega, \sigma$), and deduce the effective 
contact vertices. In more detail, we separate from the potential two 
diagrams in which the pion couples to the second nucleon and re-write 
the corresponding potentials as the product  
\begin{equation} 
V^a_2(\pi)\, \biglb( V^a_{13}+  V^a_{31}) \quad
\quad \text{with} \quad V^a_2(\pi)= - \frac{g}{2m}\, \tau_2^a\,
\bm{\sigma}_2 \cdot {\bf q}_2\,  \tilde{\Delta}_\pi({\bf q}_2)\, ,
\end{equation} 
where the factor $V^a_2(\pi)$ includes an overall factor
$i$, the $\pi NN$ vertex, and the pion propagator. Hence  $V^a_{13}
+  V^a_{31}$ is the vertex function of the effective point Lagrangian 
\begin{equation}
\mathcal{L}^{\text{point}} =  (N^\dagger \Gamma_1 N)\, 
 (N^\dagger \Gamma_3 N)\, \pi^a 
\, . 
\end{equation} 
The effective leading order Lagrangian is taken in the form
\begin{eqnarray}
 \mathcal{L}^{(1)}&=&  
 \mathcal{L}^{(1)}(\alpha)+\mathcal{L}^{(1)}(\beta)\, , \\
 \mathcal{L}^{(1)}(\alpha)&=&
 - \alpha_1 \, (N^\dagger N)(N^\dagger  \bm{\sigma} \tau^a N)
 \cdot \bm{\nabla}\pi^a - \alpha_2 \,
 (N^\dagger \bm{\sigma} N)(N^\dagger \tau^a N)\cdot \bm{\nabla}\pi^a
\nonumber\\
&& - \alpha_3\, \epsilon^{abc}\, (N^\dagger  \bm{\sigma} \tau^b N)
 \times (N^\dagger  \bm{\sigma} \tau^c N)\cdot \bm{\nabla}\pi^a \, ,
\label{3Nepela}  \\
 \mathcal{L}^{(1)}(\beta)&=&
+ \beta_1\, \pi^a\, (N^\dagger  \bm{\sigma} \tau^a N) \cdot \bm{\nabla}(N^\dagger N)
+ \beta_2 \,\pi^a \, (N^\dagger \bm{\sigma} N) \cdot \bm{\nabla} 
(N^\dagger \tau^a N)
 \nonumber \\
&& + \beta_3\, \epsilon^{abc}\, \pi^a\, \epsilon_{kij}\, 
(N^\dagger \sigma_i \tau^b N)\, \nabla_k ( N^\dagger \sigma_j \tau^c N)
 \, .
 \label{3Nepelb}
\end{eqnarray} 
It generalizes the Lagrangian considered in Ref.\ \cite{Epelbaum3N} by
including (in $\mathcal{L}^{(1)}(\beta)$) vertices with derivatives of 
nucleon fields.\footnote{In this paper we use the leading order $\pi 
NN$ coupling with the plus sign ($\mathcal{L}_{\pi NN}= +\frac{g}{2m}  
(N^\dagger  \bm{\sigma} \tau^a N) \cdot \bm{\nabla}\pi^a$), in 
agreement with H\"uber {\it et al.} \cite{HuberFBS,FriarKolck}, but 
opposite to Epelbaum  {\it et al.} \cite{Epelbaum3N}. Therefore, the 
Lagrangian (\ref{3Nepela}) has the opposite sign compared to Eq.(2.3) 
of Ref.\ \cite{Epelbaum3N}.} 

For the local part of the $\pi-\rho$ potential with a timelike 
$\rho$-exchange (\ref{pirhoPVT}) we thus obtain 
\begin{equation} 
\mathcal{L}^{\text{point}}(\pi-\rho T,\text{PV+})=
\frac{g g_\rho^2}{2m^2 m_\rho^2}\, \pi^a\, (N^\dagger \bm{\sigma}N)\, 
\bm{\nabla} (N^\dagger \tau^a N)  \, , 
\end{equation}
and comparing to (\ref{3Nepelb}) we extract
\begin{equation}
\beta_2(\pi-\rho T,\text{PV+})= \frac{g g_\rho^2}{2m^2 m_\rho^2}\,  ,
\end{equation} 
where for PS coupling there would be an additional factor $1+\kappa_\rho$. For 
the  local part of the $\pi-\rho$ KR term (\ref{pirhoPSS}) one gets 
\begin{eqnarray} 
\mathcal{L}^{\text{point}}(\pi-\rho S,\text{KR})&=& 
- \frac{g g_\rho^2 (1+\kappa_\rho)}{2m^2 m_\rho^2}\, 
\epsilon_{abc}\, \epsilon_{kij}\, \pi^a  (N^\dagger \sigma_i \tau^b N)\,
 \nabla_k ( N^\dagger \sigma_j \tau^c N) \, . \\
\beta_3(\pi-\rho S,\text{KR})&=& - 
\frac{g g_\rho^2 (1+\kappa_\rho)}{2m^2 m_\rho^2}\, .
\label{KRpoint} 
\end{eqnarray} 
For the Born $\pi-\sigma$ exchange the point low-energy Lagrangians 
corresponding to the limit of (\ref{pisigmaPV}) and (\ref{pisigmaPS}) are
\begin{eqnarray}
 \mathcal{L}^{\text{point}}(\pi-\sigma,\text{PV+})&=&
 - \frac{g g^2_\sigma}{2m^2 m_\sigma^2}\, 
\pi^a\,  (N^\dagger \tau^a \bm{\sigma}\, N)\, \cdot
 \bm{\nabla} (N^\dagger N) \, , \label{LpointsigmaPV}\\
 \beta_1(\pi-\sigma,\text{PV+})&=& - \frac{g g^2_\sigma}{2m^2 m_\sigma^2}\, ,
\\
 \mathcal{L}^{\text{point}}(\pi-\sigma,\text{PS-})&=&
 + \frac{g g^2_\sigma}{2m^2 m_\sigma^2}\, 
(N^\dagger N) \,  (N^\dagger \bm{\sigma} \tau^a\, N)\cdot 
(\bm{\nabla} \pi^a)\, , \label{LpointsigmaPS}\\
 \alpha_1(\pi-\sigma,\text{PS-})&=& - \frac{g g^2_\sigma}{2m^2 m_\sigma^2}\, .
\end{eqnarray} 
For $\pi-\omega$ exchange we obtain 
\begin{eqnarray}
 \mathcal{L}^{\text{point}}(\pi-\omega,\text{PS-})&=&
 +\frac{g g^2_\omega}{2m^2 m_\omega^2}\, 
 \pi^a \, (N^\dagger \, \bm{\sigma} \tau^a N) \cdot
 \bm{\nabla} (N^\dagger N) \, , \\
 \beta_1(\pi-\omega,\text{PS-})&=&+\frac{g g^2_\omega}{2m^2 m_\omega^2}\, .
\label{LpointomegaPV}
\end{eqnarray}
The corresponding effective interaction for the PV+
term includes an additional factor $1+\kappa_\omega \approx 1$,
which we will neglect in the following.  
For the contributions with an intermediate $N^*$
(note that the $N^*$ propagator is already used in static point
approximation) one derives
\begin{eqnarray}
 \mathcal{L}^{\text{point}}((\pi-(\sigma+\omega)),\text{N}^*)&=&
 \frac{g^*}{m(m^*-m)}\, 
 \Biglb( \frac{g_\sigma g^*_\sigma}{m_\sigma^2} -
  \frac{g_\omega g^*_\omega}{m_\omega^2}
         \Bigrb)\, 
 (N^\dagger N)(N^\dagger  \bm{\sigma} \tau^a N)
 \cdot \bm{\nabla}\pi^a 
  \, , \\
\alpha_1(\sigma+\omega,\text{N}^*)&=& 
  - \frac{g^*}{m(m^*-m)}\, 
 \Biglb( \frac{g_\sigma g^*_\sigma}{m_\sigma^2}-
  \frac{g_\omega g^*_\omega}{m_\omega^2}
         \Bigrb)\, .  
\end{eqnarray}

It is interesting that for $\pi-\sigma$ exchange the result with PS 
$\pi NN$ coupling is closer to the form suggested in Refs.\ 
\cite{Epelbaum3N,FriarKolck}. Note also that the difference between the 
Lagrangians (\ref{LpointsigmaPV}) and (\ref{LpointsigmaPS}) (we label 
it ``cont'' since it corresponds to the chiral contact term connecting 
PV and PS couplings) is equivalent to 
\begin{equation}
 \mathcal{L}^{\text{point}}(\pi-\sigma,\text{cont})= 
 \frac{g g^2_\sigma}{2m^2 m_\sigma^2}\, (N^\dagger N)\, 
 \Biglb[ (\bm{\nabla}N^\dagger)\cdot \bm{\sigma}
  \tau^a \pi^a N + 
  N^\dagger  \tau^a \pi^a \bm{\sigma}\cdot (\bm{\nabla}N) 
  \Bigrb] \, ,
\end{equation}
This interaction is exactly the same as the second term in 
(\ref{Lpicont}) (with $c \rightarrow - g g^2_\sigma/(2 m m_\sigma^2)$). 
Note that the Lagrangian (\ref{Lpicont}) follows from the re-definition 
of the pion field (\ref{redefin}), while the so-called chiral rotation 
(which transforms in lowest order from $\pi NN$ PV to PS coupling) is a 
redefinition of the nucleon field. It is argued in \cite{FriarKolck} 
that the interactions in (\ref{Lpicont}) are unnatural from the point 
of view of  ChPT and should be discarded.  The appearance of such a 
term among the effective contact interactions raises the suspicion that 
the dynamical model (conforming with chiral symmetry) with a $\sigma$ 
meson and PV $\pi NN$ coupling contains also the mechanism 
corresponding to the first  term in (\ref{Lpicont}) and that, in 
analogy to the discussion in  \cite{FriarKolck}, both should be 
discarded. Let us  also point out that the most natural way of 
introducing the $\sigma$ meson into a chirally symmetric Lagrangian is 
the linear $\sigma$ model with PS $\pi NN$ coupling (which, however, 
fixes the coupling constant $g_\sigma$). 

\subsection{Numerical results and comparison to ChPT}

In Table \ref{T:LEC} we list the numerical values for the dimensionless 
low-energy constants  
\begin{eqnarray}
 \tilde{\alpha}_i&=& 4\, f_\pi^3\, \Lambda_\chi\, \alpha_i \, , 
 \label{atil}\\
\tilde{\beta}_i&=& 4\, f_\pi^3\, \Lambda_\chi\, \beta_i \, ,  
 \label{btil}
\end{eqnarray}
where $\Lambda_\chi= 700$ MeV as in Ref.\ \cite{Epelbaum3N}. The 
constants $\tilde{\alpha}_i, \tilde{\beta}_i$ following from the 
considered heavy meson exchanges appear to be of natural size. Notice 
that -- with the exception of the potentials with $N^*$ and the 
$\pi-\sigma$ Born term (the latter only with PS $\pi NN$ coupling) -- 
all effective vertices are of the form (\ref{3Nepelb}), i.e., they 
involve derivatives of nucleon fields. In particular, we do not get any 
contributions to the effective couplings $\tilde{\alpha}_2$ and  
$\tilde{\alpha}_3$. 

There is a pronounced cancellation between $\omega$ and $\sigma$ 
exchanges in the low energy limit, both for the Born (with PV coupling) 
and Roper contributions. Although the $\omega$ and $\sigma$ terms would 
always have opposite signs, their sum is rather sensitive to  the 
poorly determined values of the coupling constants. In particular, the 
almost perfect cancellation of the respective Roper terms appears to be 
accidental. The timelike part of the $\rho$ exchange is rather small 
for PV $\pi NN$ coupling, but for PS coupling the large factor 
$1+\kappa_\rho$ increases the corresponding $\tilde{\beta}_2$ to the 
order of all the other contributions. These results are completely in 
line with our results for the corresponding potentials included in 
their full form into the Faddeev equations, i.e., the low energy limit 
changes of course the size of the individual contributions, but not 
their signs and relative magnitude. 

The chiral $3N$ forces and their effect on the $3N$ observables were 
studied in Ref.\ \cite{HuberFBS} and more recently in Ref.\ 
\cite{Epelbaum3N}. Unfortunately, the point limit of the short-range 
forces considered above cannot be compared directly to their results,    
because neither of these papers includes terms  with derivatives of 
nucleon fields in the Lagrangian. In Ref.\ \cite{Epelbaum3N} it is 
claimed that these terms can be reduced to the form of (\ref{3Nepela}) 
through integration by parts. To transform our operators into a form 
consistent with the effective Lagrangian of Ref.\ \cite{HuberFBS}, we 
make the following {\em momentum replacement}:  ${\bf q}_3 \rightarrow  
- {\bf q}_2$. It corresponds to integrating the effective Lagrangians 
with nucleon derivatives by parts (this is equivalent to the use of 
momentum conservation at corresponding vertices, ${\bf q}_3= - {\bf 
q}_2- {\bf q}_1$) and keeping only the resultant terms with a pion 
field derivative (momentum  ${\bf q}_2$), while omitting at the same 
time still another term in the Lagrangian with nucleon field 
derivatives (leading to the momentum ${\bf q}_1$).  
 
However, our numerical calculations do not justify neglecting the 
effective Lagrangians with  nucleon field derivatives. This is seen 
clearly from Table \ref{T:momrepl} where we compare the effect of 
momentum replacement in matrix elements (with the fully correlated 
wavefunction) of the original potentials, their point limits.  Since 
these numbers are quite different, we checked in addition whether the 
disagreement is due to high momentum components of the wave function by 
multiplying the operators by the overall exponential form factor of 
Ref.\ \cite{Epelbaum3N}. However, we do not see any improvement with 
respect to the agreement between  $\langle W \rangle$  and $\langle 
W_\mathrm{repl} \rangle$. 

\begin{table}[tb]
\begin{ruledtabular}
\begin{tabular}{ldddd}
 & \multicolumn{2}{c}{Paris} & \multicolumn{2}{c}{Nijmegen 93} \\
$3N$ potential $W$ & \multicolumn{1}{c}{$\langle W \rangle$} & 
\multicolumn{1}{c}{$\langle W_\mathrm{repl} \rangle$} & 
\multicolumn{1}{c}{$\langle W \rangle$} & 
\multicolumn{1}{c}{$\langle W_\mathrm{repl} \rangle$} \\
\hline
$\pi-\rho$ (KR)                                  &    0.063    &  -0.220 & 0.097 & -0.231 \\
$\pi-\rho$ (KR) point limit                 &    0.045    &  -0.223  & 0.085 & -0.250 \\
$\pi-\rho$ (KR) point limit $\times$ exp. ff.   &    0.053    &   0.064  & 0.094 & 0.154 \\
\hline
$\pi-\sigma$ (Z)                                 &    1.359    &  -1.329  & 1.632 &   -1.617\\
$\pi-\sigma$ (Z) point limit                    &    0.873    &  -0.531   & 0.921 &   -0.693\\
$\pi-\sigma$ (Z) point limit $\times$ exp. ff.  &    2.034    &  -0.015   & 2.409 &    0.186\\
\hline
$\pi-\rho$ (T)                                   &   -0.036    &  -0.012  & -0.036 & -0.016 \\
$\pi-\rho$ (T) point limit                  &   -0.050    &  -0.016   & -0.049 & -0.020\\
$\pi-\rho$ (T) point limit $\times$ exp. ff.    &   -0.027    &   0.004   & -0.024 &  0.011\\
\end{tabular}
\end{ruledtabular}
\caption{Validity of the momentum replacement prescription. Matrix 
elements (in MeV) of  $3N$ forces related by the momentum replacement 
prescription described in the text are compared. For each of the three 
considered cases also the point limit, as well as the point limit 
multiplied with an additional exponential form factor are shown. The 
matrix elements are calculated with a wave function corresponding the 
Paris $NN$ potential together with the $3N$ forces of Table 
\ref{T:expvalues}. } 
\label{T:momrepl} 
\end{table}

Nevertheless, since no other numerical estimates of LECs are currently 
available, we use the momentum replacement prescription to see how the 
LECs derived from the potentials of this paper compare with those 
deduced in Refs.\ \cite{HuberFBS,Epelbaum3N}. For the LECs the momentum 
replacement translates into $\beta_i \rightarrow \alpha_i$. 

H\"uber {\it et al.} \cite{HuberFBS} introduce two LECs $c_1$ and 
$c_2$, related to our $\tilde{\alpha}_1$ and $\tilde{\alpha}_3$ as follows
\begin{eqnarray}
c_1&=& \frac{\Lambda}{4\Lambda_\chi}\, \tilde{\alpha}_1= \frac{5}{14}\,
\tilde{\alpha}_1 \, \\
c_2&=& \frac{\Lambda}{2\Lambda_\chi}\, \tilde{\alpha}_3= \frac{10}{14}\,
\tilde{\alpha}_3 \, ,
\end{eqnarray}
with $\Lambda = 1$ GeV. They fit these constants to $A_y$ from $nd$ 
scattering, arguing that $c_1$ is fixed to about  $c_1 \sim -3$ (from 
$A_y$ at 3MeV) and $c_2 \sim 0.5 \dots 1.0$ (from $A_y$ at 10MeV). From 
the long-short range potentials considered in this paper, only the KR 
$\pi-\rho$ potential contributes in the low-energy limit to the 
constant $c_2$, yielding $c_2= - 1.55$ (see Table \ref{T:LEC}) which 
differs form the value cited above both in sign and magnitude. However, 
it should be stressed once more that the momentum replacement was used 
in deriving $c_2(KR)$ to convert the corresponding point-like vertex 
from the form (\ref{3Nepelb}) to the form (\ref{3Nepela}). The total 
$c_1$ from Table \ref{T:LEC} is $c_1 \sim 1$, again quite different 
from that of Ref.\ \cite{HuberFBS}. Here it may be worth mentioning 
that in Ref.\ \cite{HuberFBS}  the second term in the Lagrangian 
(\ref{3Nepela}) proportional to $\alpha_2$ is not considered. 

%==========================================================
%  Table of LECs

\begin{table}[bt]
\begin{center}
\begin{ruledtabular}
\begin{tabular}{lrr}
    Diagram  \quad        & $ LECs  $    &  $\quad c_i$                       \\
\hline
 $\rho-T (PV)$       & $\tilde{\beta}_2$=  0.29  &                            \\
 $\rho-T (PS)$       & $\tilde{\beta}_2$=  2.17  &                            \\
 $\rho-S (KR)$       & $\tilde{\beta}_3$= -2.17  &   $\quad c_2$(repl)= -1.55 \\
\hline
 $\omega-Born$       & $\tilde{\beta}_1$=  8.32  &   $\quad c_1$(repl)=  2.97 \\
 $\sigma-Born (PV)$  & $\tilde{\beta}_1$= -6.29  &   $\quad c_1$(repl)= -2.25 \\
 $\sigma-Born (PS)$  & $\tilde{\alpha}_1$=-6.29  &   $\quad c_1=$ -2.25  \\
\hline
 $\omega-N^*$        & $\tilde{\alpha}_1=$ 2.99  &   $\quad c_1=$ 1.07  \\
 $\sigma-N^*$        & $\tilde{\alpha}_1=$-2.26  &   $\quad c_1=$-0.81  \\ 
\end{tabular}
\end{ruledtabular}
\end{center}
\caption{Dimensionless low-energy constants defined by Eqs.\
(\ref{atil},\ref{btil}),  derived from
$\pi-\rho$, $\pi-\omega$ and $\pi-\sigma$ exchanges using the masses and
coupling constants of Table (\ref{T:3NFconstants}). The constants
$c_i$(repl) were obtained with the help of the momentum
replacement, for $\sigma-Born$ term this replacement is needed only if PV
$\pi NN$ coupling is used}
\label{T:LEC} 
\end{table}
%==================================================================

From this comparison of LECs it is at the moment hard to see whether 
the potentials used in this paper could provide a reasonable 
description of $A_y$ (and if there is therefore a reason to believe 
that we have not missed some other important short-range effects). The 
comparison of LECs is seriously hampered by the fact that the ChPT 
derivation of Refs.\ \cite{HuberFBS},\cite{Epelbaum3N} does not 
consider effective operators to which the meson-exchange potentials 
reduce to in the low-energy limit.  

In Ref.\ \cite{Epelbaum3N} it is further argued  that the Lagrangian 
(\ref{3Nepela}) reduces effectively to just one independent term when 
used with $3N$ wave functions which are antisymmetric in the spin and 
isospin subspace of the two nucleons coupled to the contact vertex. 
Hence only one effective coupling constant (called $D$) for the $\pi 
NNNN$ vertex is considered in their numerical analysis. Assuming this, 
the matrix element of the vertex function corresponding to the 
Lagrangian (\ref{3Nepela}) is reduced to the matrix element of only one 
particular combination of spin-isospin operators. Indeed, for 
\begin{equation}
 v_{13}^a = \left[ \alpha_1\, (\tau_1^a \bm{\sigma}_1+ \tau_3^a \bm{\sigma}_3)
+ \alpha_2\, (\tau_1^a \bm{\sigma}_3+ \tau_3^a \bm{\sigma}_1)+ 
2 \alpha_3\, 
(\bm{\tau}_1 \times \bm{\tau}_3)^a (\bm{\sigma}_1 \times \bm{\sigma}_3)
\right] \, , 
\end{equation}
it holds that
\begin{equation}
A_{13}\  v_{13}^a = 
\frac{\alpha_1-\alpha_2+ 4\alpha_3}{4}\, \left[
 (\bm{\tau}_1 - \bm{\tau}_3)^a\, (\bm{\sigma}_1 - \bm{\sigma}_3) -
 (\bm{\tau}_1 \times \bm{\tau}_3)^a\, (\bm{\sigma}_1 \times \bm{\sigma}_3) 
\right] \, ,
\label{antisym} 
\end{equation}
where $A_{13}= \frac{1}{2}(1- P_{13}^\tau P_{13}^\sigma )$ is the 
spin-isospin antisymmetrization operator for particles 1 and 3.  
Therefore, if only such components of $3N$ wave functions are 
important, the effect of (\ref{3Nepela}) can be represented by a single 
term. However, we would like to point out that the $3N$ wave function 
contains also components which are {\em symmetric} in spin-isospin 
space (and antisymmetric in their spatial part). For such components 
one gets instead of (\ref{antisym}) 
\begin{eqnarray}
S_{13}\  v_{13}^a &=&   \frac{\alpha_1+\alpha_2}{2}\,
 (\bm{\tau}_1 + \bm{\tau}_3)^a\, (\bm{\sigma}_1 + \bm{\sigma}_3) \,  
\nonumber\\
&& +
\frac{\alpha_1-\alpha_2- 4\alpha_3}{4}\, \left[
 (\bm{\tau}_1 - \bm{\tau}_3)^a\, (\bm{\sigma}_1 - \bm{\sigma}_3) +
 (\bm{\tau}_1 \times \bm{\tau}_3)^a\, (\bm{\sigma}_1 \times \bm{\sigma}_3) 
\right] \, ,
\label{ssym} 
\end{eqnarray}
where $S_{13}= \frac{1}{2}(1+ P_{13}^\tau P_{13}^\sigma )$ is the
spin-isospin symmetrization operator for particles 1 and 3. If these 
components are not omitted, the matrix elements of the potentials 
derived from the  $\pi NNNN$ interaction (\ref{3Nepela}) still contain 
three different combinations of the constants $\alpha_1, \alpha_2$, and 
$\alpha_3$. The components of the triton wave function with odd orbital 
angular momentum between nucleons 1 and 3 contribute about 5 per cent 
to the total norm. Even though the weight of these states in the norm 
is not large, their effect can become significant in matrix elements of 
operators with the proper tensor structure. Moreover, $NN$ P-waves are 
known to very important for the description of $A_y$ in $Nd$ 
scattering.  

If we neglect for the moment the symmetric spin-isospin wave function 
components and interactions with nucleon field derivatives, then the 
effect of the $\alpha_i$ terms is reduced to the calculation of the 
matrix element of (\ref{antisym}). In this approximation, the 
dimensionless constant $c_D$ of Ref.\ \cite{Epelbaum3N} is given in 
terms of $\tilde{\alpha}_i$ as 
\begin{equation}
 c_D= - \tilde{\alpha}_1+ \tilde{\alpha}_2- 4 \tilde{\alpha}_3  \, .
 \label{cd}
\end{equation}
From Table \ref{T:LEC} we get $c_D \sim 6 \dots 8$, depending on the 
type of the $\pi NN$ coupling. This numerical value of $c_D$ is larger 
than $c_D \sim 1.8 \dots 3.6$ obtained in Ref.\ \cite{Epelbaum3N}. In 
fact, the KR contribution dominates our value of $c_D$, due to the 
factor 4 in the last term of (\ref{cd}). As in the comparisons above to 
Ref.\ \cite{HuberFBS}, we had to compare quite different effective 
Lagrangians and the approximations made in establishing the 
correspondence between the effective constants appear to be too crude 
to allow drawing clear conclusions. 

Finally we note that the efforts of determining the LECs in 
\cite{HuberFBS} may have resulted in a better description of $A_y$ with 
different constants, had the additional freedom in the parameters of 
the TM $\pi-\pi$ force due to the experimental uncertainties discussed 
in Sec.\ \ref{S:pipi} been taken into account.

\section{Conclusions}

The conclusions of this paper can be summarized in three points:

1. When considering the effect of the TM $\pi-\pi$ exchange 3$N$ force 
on the triton binding energy, only the variation of the cut-off 
parameter, which is often adjusted to reproduce the experimental value, 
is studied in the literature. 

However, the $\pi-\pi$ force contains the $\pi$N scattering amplitude 
as a building block, which at low pion momenta has a model independent 
form parameterized by three constants which have to be extracted from 
experiment. In contrast to the cut-off parameter, each of these three 
constants multiply different spin-isospin operators, and therefore act 
differently on the various channels of the wavefunction. We calculate 
the propagation of the experimental errors of the $\pi N$ data, which 
are used to extract the off-shell TM $\pi N$ scattering amplitude, into 
the $3N$ force parameters. As a consequence, the triton binding energy 
calculated with the TM $\pi-\pi$ potential has an uncertainty of about 
$\pm 0.4$ MeV, which is almost entirely due to the experimental errors 
in the nucleon sigma term. 

2. The long-short range $\pi-\sigma$ and $\pi-\omega$ exchange $3N$ 
forces individually have large effects on the triton binding, but two 
kinds of cancellations determine their net effect. We find contributions 
of opposite sign and comparable or very close in magnitude from 
$Z$-graph and $N^*$ excitation, respectively, in the $\pi-\sigma$ and 
$\pi-\omega$  $3N$ forces. Also $Z$-graph and $N^*$ potentials cancel 
in part for each of these long-short range forces separately. 

The extent of the cancellation is controlled decisively by the ratio 
between the $\sigma N^* (1440) N$ and $\omega N^* (1440) N$ coupling 
constants. If this ratio is taken to be the same as for the $\sigma NN$  
and $\omega NN$ coupling constants, as suggested by the naive 
constituent quark model, the attraction of the $\omega$ exchange, 
originated by the Born terms of relativistic order, dominates.  

The change in binding energy caused by these $3N$ potentials cannot be 
calculated perturbatively, since their inclusion in the Hamiltonian 
changes the wavefunction significantly. For instance, we find that, 
although the net effect of $\pi-\sigma$ and $\pi-\omega$ exchange $3N$ 
forces on the binding energy is small, the resulting wavefunction 
yields rather different matrix elements of individual interactions from 
the ones calculated without those terms in the Hamiltonian. Therefore, 
it would be very interesting to study these variations in the 
wavefunction, for instance in electron scattering on the $3N$ bound 
state. Furthermore, these potentials may strongly influence low-energy 
$Nd$ scattering observables. 

In the case of PS $\pi NN$ coupling, both  $Z$-diagrams are attractive, 
resulting in a strong overbinding of the triton. In the future, these 
$Z$-diagrams should be complemented by additional potentials derived 
from chiral contact $\pi NN \sigma$ and  $\pi NN \omega$ vertices.

3. We extracted the LECs  from the low-energy limit of the long-short 
range meson exchange potentials and compared with the ChPT results of 
Refs.\ \cite{HuberFBS,Epelbaum3N}. The constants we obtain are of 
natural size. Unfortunately, a direct comparison with the ChPT results 
is not possible since we arrive at contact vertices of different 
structure. To translate them into the ChPT form certain approximations 
are necessary, the validity of which we found to be unfounded in our 
numerical evaluation. It does not come as a surprise that our LECs do 
not agree with those of Refs.\ \cite{HuberFBS,Epelbaum3N}. Efforts 
should be joined from the two sides (meson-exchange description and 
ChPT methods) to investigate this question further.

\begin{acknowledgments} 
J.A. was supported by the grants GA CR 202/00/1669, GA CR 202/03/0210 
and by the Lisbon project. He would like to thank his Portuguese 
colleagues for the warm hospitability during his stay in Lisbon. M.T.P. 
and A.S. were supported by FCT and FEDER under grants 
CERN/FIS/43709/2001 and POCTI/FNU/40834/2001. 
\end{acknowledgments}

\appendix
\section{Contributions of nucleon Born diagrams}
\label{A:Born}

We will give here some details of the derivation of nucleon Born 
diagram contributions to the $3N$ force from the exchange of one pion 
and one heavy meson ($B$), since we experienced some misunderstandings 
and many questions when discussing their origin. Our treatment is a 
condensed version of the technique developed in a number of papers on 
e.m.\ meson exchange currents \cite{ATA,Cebaf-talk,Smatrix}. We pay 
special attention to the dependence of our results on the type of the 
$\pi NN$ vertex. 

The generic Feynman amplitude $\mathcal{W}(1)$ corresponding to
Fig.\ \ref{F:short-longandlong-long}a reads:
\begin{equation}
\mathcal{W}(1)=\mathcal{F}^a\,  \mathcal{D}^a + 
2 \leftrightarrow 3 \, ,  
\label{Wfeyn}
\end{equation}
where we gather all factor connected with the ``active'' first nucleon into
the amplitude $\mathcal{F}$ and the rest into the amplitude  $\mathcal{D}$
\begin{eqnarray}
\mathcal{D}^a&=& 
 -\, \bar{u}({\bf p}_2^\prime) \Gamma^a(\pi, q_2) u({\bf p}_2)\, 
 \bar{u}({\bf p}_3^\prime) \Gamma (B, q_3) u({\bf p}_3)\,
 \Delta_\pi(q_2)\,  \Delta_B(q_3)\, , \label{Dfeyn}\\
\mathcal{F}^a&=& 
\bar{u}({\bf p}_1^\prime) \Biglb[ 
\Gamma^a(\pi ,-q_2)\, S(P')\, \Gamma (B ,-q_3)+ 
\Gamma (B ,-q_3)\, S(P)\, \Gamma^a(\pi ,-q_2) \Bigrb] u({\bf p}_1)\, ,
\label{Ffeyn}
\end{eqnarray}
where $S(p)= (i \gamma_\mu p_\mu+ m)^{-1}$ and $\Delta_B(q)= (m_B^2+
q^2)^{-1}$ (the Pauli metric is used), $P= p'_1+ q_3= p_1- q_2,\, P'= p_1-
q_3= p'_1+ q_2$ with $q_i= p'_i- p_i$, and we suppress (unless we specify $B$)
additional isospin and Lorentz indices,  that would appear for non-scalar and
charged meson $B$.   The pion ($\pi NN$)  vertex functions
(corresponding to $i \mathcal{L}$) are  
\begin{eqnarray}
\Gamma^a(\pi\text{-PS})&=& -g \gamma_5 \tau^a \, , \\
\Gamma^a(\pi\text{-PV},q)&=& i 
\frac{g}{2m} q_\mu \gamma_\mu \gamma_5 \tau^a \, ,
\end{eqnarray}
where $q= p'-p$ is the momentum of the pion entering the vertex (the 
signs of our vertex functions differ from those of Ref.\ \cite{ATA}, we 
adopted the current convention to agree with that usually used in ChPT 
calculations). The vertex functions $\Gamma (B,q)$ for the heavy mesons 
$B$ will be specified later. Note, that from the identities 
\begin{eqnarray}
 \bar{u}({\bf p}_1^\prime)  
 \Gamma^a(\pi\text{-PV},-q_2)\, S(P') &=& \bar{u}({\bf p}_1^\prime)  
 \Biglb[ \Gamma^a(\pi\text{-PS})\, S(P')+ \frac{g}{2m}\gamma_5 \tau_1^a
 \Bigrb]  \, , \\
S(P)\, \Gamma^a(\pi\text{-PV},-q_2) u({\bf p}_1)  &=&  \Biglb[ S(P)\,
\Gamma^a(\pi\text{-PS}) + \frac{g}{2m}\gamma_5 \tau_1^a \Bigrb] u({\bf p}_1)
\, , 
\end{eqnarray}
it follows 
\begin{eqnarray}
\mathcal{F}^a_{\text{PV}} &=& \mathcal{F}^a_{\text{PS}}+ 
   \mathcal{F}^a_{\text{cont}}
\, , \\
\mathcal{F}^a_{\text{cont}}&=& \frac{g}{2m}\,
\bar{u}({\bf p}_1^\prime)\, \Biglb\{ \gamma_5 \tau_1^a, \, \Gamma (B ,-q_3)
\Bigrb\}_+ \, u({\bf p}_1) \, .
\label{Fequiv}
\end{eqnarray}
The ``contact'' amplitude arises from the usual chiral contact
interaction, which appears in the chiral rotation from the PV to
PV $\pi NN$ coupling. 

To get the quantum-mechanical $3N$ potential from the Feynman amplitude
(\ref{Wfeyn}), it is necessary to subtract the part which is in the 
quantum mechanical description already included in the $T$ matrix in 
the iterations of the $NN$ potential. We split the nucleon propagator 
into its positive and negative energy parts $S(p)= S^+(p)+ S^-(p)$: 
\begin{equation}
 S(p)= \frac{m- i\gamma_\mu p_\mu}{p^2+m^2}= 
 \frac{m+E\gamma_4- i\bm{\gamma}\cdot{\bf p}}{2E(E-p_0)}+ 
 \frac{m-E\gamma_4- i\bm{\gamma}\cdot{\bf p}}{2E(E+p_0)}=  
 \frac{u({\bf p})\bar{u}({\bf p})}{E- p_0} -
 \frac{v(-{\bf p})\bar{v}(-{\bf p})}{E+ p_0} \, ,
\end{equation}
(where $E^2= {\bf p}^2+ m^2$) and define the corresponding amplitudes
$\mathcal{W}^\pm(1)$ and $\mathcal{F}^\pm$. The spinors $u$ and $v$ are
given by:
\begin{equation}
u({\bf p})= \sqrt{\frac{E+m}{2E}}\, \left(
\begin{array}{c}
1 \\
 \frac{\bm{\sigma}\cdot {\bf p}}{E+m}
\end{array} \right) \, ;
\quad  \quad
v(-{\bf p})= \sqrt{\frac{E+m}{2E}}\, \left(
\begin{array}{c}
 - \frac{\bm{\sigma}\cdot {\bf p}}{E+m} \\
 1
\end{array} \right) 
\, .
\end{equation}
For the calculations of this paper it is sufficient to keep only the
leading order in $p/m$ and replace $E \rightarrow m$.

The true ``pair'' (or ``Z-diagram'') contributions to the $3N$
potential are defined by the  straightforward non-relativistic 
reduction (i.e.\ the decomposition of the  spinor matrix elements in 
powers of $p/m$  keeping only leading order terms) of the 
$\mathcal{W}^-(1)$. In the order considered, this contribution is 
non-zero only for the PS $\pi NN$ coupling for any meson exchange. For 
this coupling (and again, at the given order in $p/m$) 
$\mathcal{W}^+(1)$ with positive energy nucleon in the intermediate 
state corresponds to the iteration of the lowest order non-relativistic 
OPEP potential (which is independent of the type of $\pi NN$  
coupling). On the other hand, for the PV $\pi NN$ coupling, the pair 
diagrams $\mathcal{W}^-(1)$ do not contribute. But the positive energy 
part $\mathcal{W}^+(1)$ does not exactly equal to the iteration of the 
non-relativistic OPEP, since in the Feynman amplitude the vertex 
function is off-mass-shell ($P^2 \neq m^2 \neq P'^2$) while in the 
iteration of the OPEP this potential is off-energy-shell. One can 
re-arrange the energies in $q_{20}$ entering the PV vertex function in 
(\ref{Wfeyn},\ref{Ffeyn}) identically 
\begin{eqnarray}
q_{20}&=& P'_0- E(p'_1)= (E(P')- E(p'_1))- (E(P')-P'_0) \, , \\
q_{20}&=& E(p_1)- P_0  = (E(p_1)- E(P))+ (E(P)-P_0) \, . 
\end{eqnarray}
The first energy differences on the r.h.s.\ put the PV vertex on its 
mass shell. The corresponding part of $\mathcal{W}^+(1)$ is then again 
identified with the iteration of the non-relativistic OPEP and it is 
identical to the full $\mathcal{W}^+(1)$ for the PS coupling. But the 
second terms on $q_{20}$ cancel the denominator of $S^+(P')$ or 
$S^+(P)$ and give rise to a contribution to the quantum mechanical 
potential $W^+(1)$. 

To sum it up, the relevant contributions to the $3N$ potential are
\begin{eqnarray}
W(1)&=& F^a\,  D^a + 2 \leftrightarrow 3 \, , \label{Wqm}\\
D^a& \approx & 
 \frac{g}{2m} \, \tau_2^a  (\bm{\sigma}_2 \cdot {\bf q}_2) \, \,
 \bar{u}({\bf p}_3^\prime) \Gamma (B, q_3) u({\bf p}_3)\,
 \Delta_\pi(q_2)\,  \Delta_B(q_3)\, , \label{Dqm}\\
F^{a,-}_{\text{PS}} & \approx & \frac{g}{2m}
\Biglb[ \bar{u}({\bf p}_1^\prime) \gamma_5 \tau_1^a v(-{\bf P'})\,
         \bar{v}(-{\bf P'})\Gamma (B ,-q_3) u({\bf p}_1)+ \nonumber\\
 && \quad  \quad \ \quad \bar{u}({\bf p}_1^\prime) \Gamma (B ,-q_3)v(-{\bf P})\,
    \bar{v}(-{\bf P})\gamma_5 \tau_1^a u({\bf p}_1)\Bigrb] 
     \, , \label{FPSqm}\\
F^{a,+}_{\text{PV}} & \approx &      - \frac{g}{2m}
\Biglb[ \bar{u}({\bf p}_1^\prime) \gamma_4 \gamma_5 \tau_1^a u({\bf P'})\,
         \bar{u}({\bf P'})\Gamma (B ,-q_3) u({\bf p}_1)- \nonumber\\
 && \quad  \quad \ \quad \bar{u}({\bf p}_1^\prime) \Gamma (B ,-q_3)u({\bf P})\,
    \bar{u}({\bf P})\gamma_4 \gamma_5 \tau_1^a u({\bf p}_1)\Bigrb] 
     \, , \label{FPVqm}\\
F^a_{\text{cont}} & \approx & \frac{g}{2m}\,
\bar{u}({\bf p}_1^\prime)\, \Biglb\{ \gamma_5 \tau_1^a, \, \Gamma (B ,-q_3)
\Bigrb\}_+ \, u({\bf p}_1) \, .
\end{eqnarray}
where $\approx$ stands for the non-relativistic reduction. Since only 
$F^{a,-}$ survives for PS coupling and only $F^{a,+}$ contributes for
the PV one (and since the iterations of non-relativistic OPEP are 
identical in both cases), one gets from (\ref{Fequiv}) 
\begin{equation}
F^{a,+}_{\text{PV}}= F^{a,-}_{\text{PS}}+ F^a_{\text{cont}} \, ,
\label{Fident}
\end{equation}
and therefore 
\begin{equation}
W^{+}_{\text{PV}}= W^{-}_{\text{PS}}+ W_{\text{cont}} \, ,
\end{equation}
for all heavy meson exchanges considered.

It remains to list the vertex functions for the $\sigma, \omega$ and 
$\rho$ mesons and the results of the non-relativistic reduction. For 
the scalar isoscalar $\sigma$ meson one gets 
\begin{eqnarray}
\Gamma (\sigma)&=& i g_\sigma \, , \\
D^a(\sigma)&=& +i \frac{g g_\sigma}{2m}\, \tau^a_2 
(\bm{\sigma}_2 \cdot {\bf q}_2)\, 
\Delta_\pi(q_2) \Delta_\sigma(q_3) \, , \\
F^{a,+}_{\text{PV}}(\sigma)&=& -i \frac{g g_\sigma}{2m^2}\, \tau^a_1 
(\bm{\sigma}_1 \cdot {\bf q}_3)\, , \\
F^{a,-}_{\text{PS}}(\sigma)&=& +i \frac{g g_\sigma}{2m^2}\, \tau^a_1 
(\bm{\sigma}_1 \cdot {\bf q}_2)\, , \\
F^{a}_{\text{cont}}(\sigma)&=& +i \frac{g g_\sigma}{2m^2}\, \tau^a_1 
(\bm{\sigma}_1 \cdot {\bf q}_1)\, , 
\end{eqnarray}
where (\ref{Fident}) can be verified with the help of ${\bf q}_1+ 
\bf{q}_2 + \bf{q}_3= 0$. The corresponding potentials are just the 
products of $D$ and $F$ factors, they are listed in the main body of 
the paper. 

For the isoscalar vector meson $\omega$ only the timelike part
$\mu=4$ of the vertices contribute up to the order considered
\begin{eqnarray}
\Gamma_\mu (\omega,q)&=&  g_\omega \Biglb[ \gamma_\mu- 
\frac{\kappa_\omega}{2m} \sigma_{\mu \nu} q_\nu \Bigrb] \, , \\
D^a_4(\omega)&=&  \frac{g g_\omega}{2m}\, \tau^a_2 
(\bm{\sigma}_2 \cdot {\bf q}_2)\, 
\Delta_\pi(q_2) \Delta_\omega(q_3) \, , \\
F^{a,+}_{\text{PV},4}(\omega)&=& 
 - \frac{g g_\omega}{2m^2}\, \tau^a_1 
(\bm{\sigma}_1 \cdot {\bf q}_3)\, , \\
F^{a,-}_{\text{PS},4}(\omega)&=&- \frac{g g_\omega (1+\kappa_\omega)}{2m^2}\, 
\tau^a_1 (\bm{\sigma}_1 \cdot {\bf q}_3)\, , \\
F^{a}_{\text{cont}}(\omega)&=& + \frac{g g_\omega \kappa_\omega}{2m^2}\, 
\tau^a_1 (\bm{\sigma}_1 \cdot {\bf q}_3)  \, . 
\end{eqnarray}

Finally, for the isovector vector meson $\rho$ the vertex function
reads
\begin{equation}
\Gamma^b_\mu (\rho,q)=  g_\rho \Biglb[ \gamma_\mu- 
\frac{\kappa_\rho}{2m} \sigma_{\mu \nu} q_\nu \Bigrb]\, \tau^b \, . \\
\end{equation}
For the timelike component of the $\rho$ field  ($\mu=4$) one gets
\begin{eqnarray}
D^{ab}_4(\rho)&=&  \frac{g g_\rho}{2m}\, \tau^a_2 \tau^b_3 
(\bm{\sigma}_2 \cdot {\bf q}_2)\, 
\Delta_\pi(q_2) \Delta_\rho(q_3) \, , \\
F^{ab,+}_{\text{PV},4}(\rho)&=& 
 - \frac{g g_\rho}{2m^2}\, \Biglb[ \delta_{ab} (\bm{\sigma}_1 \cdot {\bf q}_3)\,
 - i \epsilon_{abc} \tau^c_1 (\bm{\sigma}_1 \cdot {\bf Q}_1) \Bigrb]
 , \\
F^{ab,-}_{\text{PS},4}(\rho)&=& 
 - \frac{g g_\rho (1+ \kappa_\rho)}{2m^2}\, \delta_{ab} 
(\bm{\sigma}_1 \cdot {\bf q}_3)\, , \\
F^{ab}_{\text{cont},4}(\rho)&=& + \frac{g g_\rho}{2m^2}
\Biglb[ \delta_{ab} \kappa_\rho (\bm{\sigma}_1 \cdot {\bf q}_3)\,
 + i \epsilon_{abc} \tau^c_1 (\bm{\sigma}_1 \cdot {\bf Q}_1) \Bigrb]\, \, , 
\end{eqnarray}
where $\bf{Q}_i= \bf{p}'_i+ \bf{p}_i$. For the spacelike component
\begin{eqnarray}
{\bf D}^{ab}(\rho)&=& - i \frac{g g_\rho}{4m^2}\, \tau^a_2 \tau^b_3 
(\bm{\sigma}_2 \cdot {\bf q}_2)\, 
\Biglb[ {\bf Q}_3 + (1+ \kappa_\rho) i {\bm \sigma}_3 \times {\bf q}_3 \Bigrb] 
\Delta_\pi(q_2) \Delta_\rho(q_3) \, , \\
{\bf F}^{ab,+}_{\text{PV}}(\rho)&=& 0 \, , \\
{\bf F}^{ab,-}_{\text{PS}}(\rho)&=& - {\bf F}^{ab}_{\text{cont}}(\rho)=
 - \frac{g g_\rho}{m}\, \epsilon_{abc} \tau^c_1\, {\bm \sigma}_1   \, .  
\end{eqnarray}

In the main body of the paper we denote the contributions due to the
forth component of vector fields by subscript $T$ and those from the
exchanges with $\mu= 1,2,3$ by the subscript $S$.

%\newpage
\section{$\sigma$-meson exchange in $3N$ potentials}
\label{A:sigma} The Bonn potentials use different parameters (masses, 
coupling constants and cut-off parameters) of their ``$\sigma$''-meson 
in $NN$ channels with isospin $I=0$ and $I=1$. Let us define  
\begin{equation}
v(\sigma(I),{\bf q}^2) = \frac{g_{\sigma(I)}^2}{4\pi} 
\frac{F^2_{\sigma(I) NN}(\Lambda_{\sigma(I)NN},{\bf q}^2)}
{m^2_{\sigma(I)} + {\bf q}^2} \, , 
\end{equation}
where $g_{\sigma(I)}, \Lambda_{\sigma(I)NN}, m^2_{\sigma(I)}$ are the 
Bonn $\sigma(I)$-exchange parameters in the respective $NN$ isospin 
channel. The dominant central part of the Bonn $\sigma$-exchange 
potential is then given by 
\begin{eqnarray}
v_C({\bf q}^2)= v(\sigma(0),{\bf q}^2)\, P_0 &+& 
                v(\sigma(1),{\bf q}^2)\, P_1\quad ,\\ 
 P_0= \frac{1}{4} (1+ \bm{\tau}_1 \cdot \bm{\tau}_2 )\, , &&
 P_1= \frac{1}{4} (3- \bm{\tau}_1 \cdot \bm{\tau}_2 ) \, ,
\end{eqnarray}
which can be re-written in a form {\em similar} to the central 
potentials originated from exchanges of two scalar-isoscalar and two 
scalar-isovector particles: 
\begin{equation}
v_C({\bf q}^2)= \frac{1}{4} \left[ v(\sigma(0),{\bf q}^2)+ 
3\, v(\sigma(1),{\bf q}^2) \right]
+ \bm{\tau}_1 \cdot \bm{\tau}_2 \,
  \frac{1}{4} \left[ v(\sigma(0),{\bf q}^2)- v(\sigma(1),{\bf q}^2) \right] \, .
  \label{isosep}
\end{equation}
This does not mean that the Bonn $\sigma$-exchange is {\em equivalent} 
to the true exchange of four scalar particles, since: 1) the last term 
has the sign opposite to the sign of a real scalar exchange; 2) all 
four of these exchange have to act at once, i.e., they cannot be 
separated by an exchange of another meson (in iterations of the 
Lippmann-Schwinger equation). 

\begin{figure}
\epsfig{file=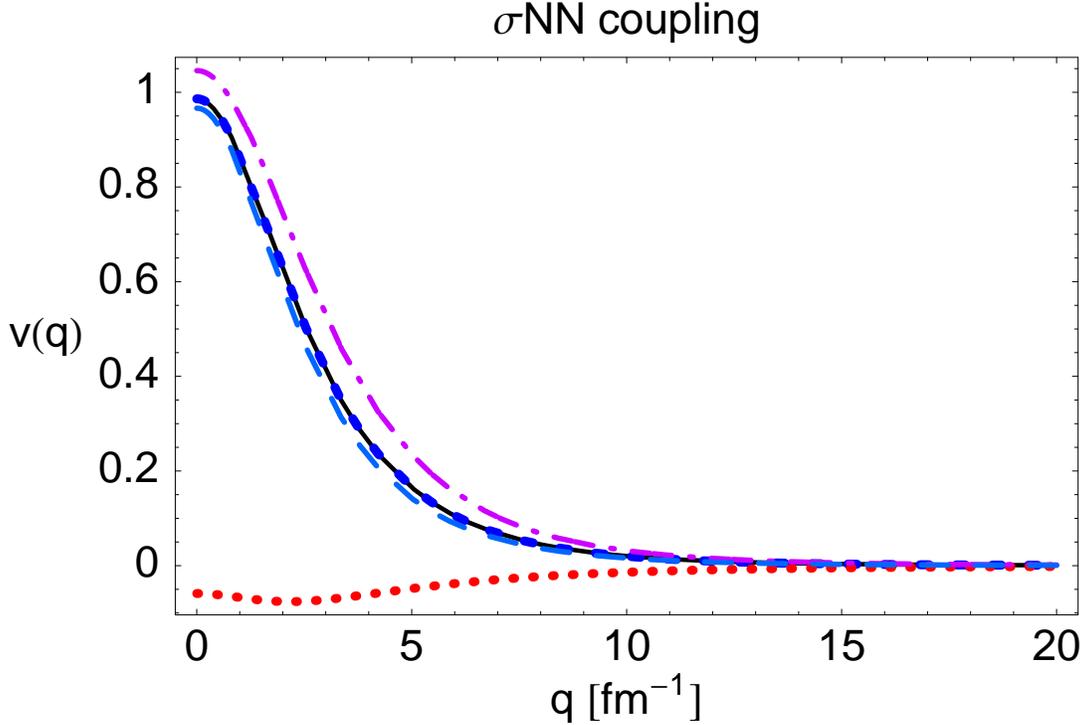,width=10cm,angle=-90} 
\caption{The fit (solid line) to the central part of the isoscalar Bonn 
potential (short-dashed line; practically coincides with the solid line 
on the graph), the first term of Eq.\ (\ref{isosep}). For comparison, 
also $v(\sigma(0),{\bf q}^2)$ (dash-dotted line), $v(\sigma(1),{\bf 
q}^2)$ (dashed line) and their isovector combination (the second term 
in Eq.\ (\ref{isosep}); dotted line) are shown.} 
\label{F:sigmafit} 
\end{figure}

To have a simple prescription for the  $\sigma$-exchange in our $3N$ 
potentials, we fitted the first term in Eq.\ (\ref{isosep}) by the 
function $v(\sigma,{\bf q}^2)$ dependent on parameters of single true 
isoscalar $\sigma$-exchange $g_\sigma, \Lambda_{\sigma NN}, 
m^2_\sigma$. The result of the fit to the Bonn B potential gives the 
values listed in Table \ref{T:3NFconstants}, and the quality of the fit 
is shown on Fig.\ \ref{F:sigmafit}. The fitted parameters are between 
the Bonn values for $NN\ I= 0,1$ channels. The second 
(``isovector-exchange'') component of Eq.\ (\ref{isosep}) is much 
smaller in absolute value and even negative for $q \leq 10 fm^{-1}$ 
(see Fig.\ \ref{F:sigmafit}), which precludes an approximation by the 
exchange of scalar particle(s). If we nevertheless include it into the 
$3N$ potential, it gives a considerably smaller contribution to the 
triton binding energy. Therefore, we neglected it in the calculations 
of this paper. 

\newpage 
 
%-------------------------------------------------------------
%\begin{thebibliography}

%\end{thebibliography}


\begin{references}
%1
\bibitem{CD-Bonn} R. Machleidt, Phys. Rev. C {\bf 63}, 024001 (2001).
%2
\bibitem{Nijmegen} V.G.J. Stocks, R.A.M. Klomp, C.P.F. Terheggen, 
          J.J. de Swart, Phys. Rev. C {\bf 49}, 2950 (1994).
%3
\bibitem{Bonn}
  R. Machleidt, K. Holinde, and Ch. Elster, Phys. Rep. {\bf 149}, 1 (1987).
%4
\bibitem{Beane}
 S.R. Beane, P.F. Bedaque, M.J. Savage, and U. van Kolck, 
 Nucl. Phys. {\bf A700}, 377 (2002).  
%5
\bibitem{Epelbaum}
  E. Epelbaum, W. Gl\"ockle, Ulf-G. Mei\ss ner, Nucl. Phys. {\bf A637}, 107 (1998).
%6
\bibitem{Epelbaumnew}  
  E. Epelbaum, W. Gl\"ockle, Ulf-G. Mei\ss ner, Nucl. Phys. {\bf A671}, 295 (2000);\\
 E. Epelbaum, A. Nogga, W. Gl\"ockle, H. Kamada, Ulf-G. Mei\ss ner, H. Wita\l a,
 Eur. Jour. Phys. {\bf A15}, 543 (2002).
%7  
\bibitem{Machleidt} D.R. Entem and R. Machleidt,  
  Phys. Rev. C {\bf 65},  064005 (2002); 
  Phys. Rev. C {\bf 66}, 014002 (2002).   
%8
\bibitem{saturation}
 E. Epelbaum, Ulf-G. Mei\ss ner, W. Gl\"ockle, Ch. Elster, 
 Phys. Rev. C {\bf 65}, 044001 (2002). 
%9
\bibitem{HuberFBS} D. H\"uber, J.L. Friar, A. Nogga, H. Wita\l a,
   U. van Kolck,  Few-Body Syst.\ {\bf 30}, 95 (2001).
%10
\bibitem{Epelbaum3N}
 E. Epelbaum, A. Nogga, W. Gl\"ockle, H. Kamada, Ulf-G. Mei\ss ner, H. Wita\l a,
Phys. Rev. C {\bf 66}, 064001 (2002).
%11
\bibitem{TM}
  S.A. Coon, M.D. Scadron, and B.R. Barrett, 
             Nucl. Phys. {\bf A242}, 467 (1975);\\
  S.A. Coon, M.D. Scadron, P.C. McNamee, B.R. Barrett, D.W.E. Blatt, 
  B.H.J. McKellar, Nucl. Phys. {\bf A317}, 242 (1979).
%12
\bibitem{TM1}
  S.A. Coon and W. Gl\"ockle, Phys. Rev. C {\bf 23}, 1790 (1981).  
%13
\bibitem{Ellis}
   R.G. Ellis, S.A. Coon, and B.H.J. McKellar, 
     Nucl. Phys. {\bf A438}, 631 (1985).
%14  
\bibitem{CPR} S.A. Coon, M. T. Pe\~na, and D.O. Riska, 
        Phys. Rev. C {\bf 52}, 2925 (1995).
%15 
\bibitem{FriarKolck} J.L. Friar, D. H\"uber, U. van Kolck,
            Phys. Rev. C {\bf 59}, 53 (1999).
%16       
\bibitem{GBrown} G.E. Brown, 
     Comments on Nucl. and Part. Phys. {\bf 4}, 140 (1970);\\
  G.E. Brown and J.W. Durso, Phys. Lett. {\bf 35B}, 120 (1971).
%17
\bibitem{Robilotta} H.T. Coelho, T.K. Das, and M.R. Robilotta, 
         Phys. Rev. C {\bf 28}, 1812 (1983).
%18
\bibitem{Weinberg} S. Weinberg, Physica {\bf 96A}, 327 (1979).
%19
\bibitem{Kolck} U. van Kolck, Phys. Rev. C {\bf 49}, 2932 (1994); \\
   C. Ordon\^ez, L. Ray, and U. van Kolck, Phys. Rev. C {\bf 53} (1996) 2086. 
%20
\bibitem{CoonHan}  S.A. Coon, H.K. Han, 
            Few Body Syst. {\bf 30}, 131 (2001). 
%21
\bibitem{FriarHuber} D. H\"uber and J.L. Friar, 
           Phys. Rev. C {\bf 58}, 674 (1998).
%22
\bibitem{Kievsky} A. Kievsky, 
           Phys. Rev. C {\bf 60}, 034001 (1999). 
%23
\bibitem{Meissner} E. Epelbaum, H. Kamada, A. Nogga, H. Wita\l a,
     W. Gl\"ockle, Ulf-G. Mei\ss ner, Phys. Rev. Lett. {\bf 86}, 4787 (2001). 
%24
\bibitem{Canton} L. Canton, W. Schadow, 
          Phys. Rev. C {\bf 62}, 044005 (2002);\\
      L. Canton, W. Schadow, J. Haidenbauer, 
         Eur. Phys. J. {\bf A14}, 225 (2002).  
%25 
\bibitem{Pieper}  S.C. Pieper, V.R. Pandharipande, R.B. Wiringa,
   J. Carlson, Phys. Rev. C {\bf 64}, 014001 (2001).
%26
\bibitem{Urbana} B.S. Pudliner, V.R. Pandharipande, J. Carlson, S.C. Pieper, 
     and R.B. Wiringa, Phys. Rev. C {\bf 56}, 1720 (1997).
%27
\bibitem{matter}     
   H. Heiselberg, V. Pandharipande, 
      Ann. Rev. Nucl. Part. Sci. {\bf 50}, 481 (2000). 
%28
\bibitem{piBorn} 
   S.A. Coon, W. Gl\"ockle, Phys. Rev. C {\bf 23}, 1796 (1981);\\
   S.A. Coon, J.L. Friar, Phys.Rev. C {\bf 34}, 1060  (1986).
%29    
\bibitem{PenaCoon} S.A. Coon, M.T. Pe\~na, 
      Phys. Rev. C {\bf 48}, 2559 (1993).
%30
\bibitem{Stadler}  A.~Stadler, J.~Adam Jr.\ , H.~Henning, and P.~U.~Sauer,
       Phys.\ Rev.\ C {\bf 51}, 2896 (1995). 
%31
\bibitem{Smejkal} J. Smejkal,  E. Truhl\'{\i}k, H. G\"oller,   
 Nucl. Phys. {\bf A624}, 655 (1997), and references therein.
%32
\bibitem{Soyeur} M.~Soyeur, Nucl. Phys. {\bf A671}, 532 (2000).
%33
\bibitem{Oset} S.~Hirenzaki, P.~Fern\'andez de C\'ordoba,
   E.~Oset,  Phys. Rev. C {\bf 53}, 277 (1996).
%34
\bibitem{Stadler91} A. Stadler, W. Gl\"ockle, and P.U. Sauer, Phys. Rev. C {\bf 44}, 2319
(1991).
%35
\bibitem{Cebaf-talk} J. Adam, Jr.,
 ``Proceedings of XIVth International Conference on Few Body
  Problems in Physics'', May 26-31, 1994, Williamsburg, Virginia
edited by Franz Gross, AIP {\bf 334}, 192 (1994).
%36
\bibitem{Smatrix} M. Chemtob, M. Rho, Nucl. Phys. {\bf A163}, 1 (1971);\\
    D.O. Riska, Prog. Part. Nucl. Phys. {\bf 11}, 199 (1984).
%37
\bibitem{ATA} J. Adam, Jr.,  E. Truhl\'{\i}k, D. Adamov\'a, 
     Nucl. Phys. {\bf A492}, 556, (1989).
%38
\bibitem{Reid} R. V. Reid, Ann. Phys. (N.Y.) {\bf 50}, 411 (1968).
%39
\bibitem{Paris} M. Lacombe, B. Loiseau, J. M. Richard, R. Vinh Mau,
                J. C\^ot\'e, P. Pir\`es, and R. de Tourreil,  
                Phys. Rev. C {\bf 21}, 861 (1980).
%40
\bibitem{BonnB} R. Machleidt, Adv. Nucl. Phys. {\bf 19}, 189 (1989).
% 
\end{references}
\end{document}